%% file: submit_main.tex
\documentclass{SCIS2025}
\usepackage{comment}

\begin{document}
%\oa
%%%%%%%%%%%%%%%%%%%%%%%%%%%%%%%%%%%%%%%%%%%%%%%%%%%%%%%
%%% Authors do not modify the information below
%%% ×÷Õß²»ÐèÒªÐÞ¸Ä´Ë´¦ÐÅÏ¢
\ArticleType{POSITION PAPER}
%\SpecialTopic{}
%\luntan
\Year{2025}
\Month{}
\Vol{}
\No{}
\DOI{}
\ArtNo{}
\ReceiveDate{}
\ReviseDate{}
\AcceptDate{}
\OnlineDate{}
\AuthorMark{}
\AuthorCitation{}
%%%%%%%%%%%%%%%%%%%%%%%%%%%%%%%%%%%%%%%%%%%%%%%%%%%%%%%

%%% title: ±êÌâ
%%%   \title{title}{title for citation}
\title{Large Processor Chip Model}{Large Processor Chip Model}

%%% Corresponding author: Í¨ÐÅ×÷Õß
%%%   \author[number]{Full name}{{email@xxx.com}}
%%% General author: Ò»°ã×÷Õß
%%%   \author[number]{Full name}{}
%\author[]{Yunji Chen}{}
%\author[]{Kaiyan Chang}{}
%\author[]{Mingzhi Chen}{}
%\author[]{Zhirong Chen}{}
%\author[]{Dongrui Fan}{}
%\author[]{Junfeng Gong}{}
%\author[]{\\Nan Guo}{}
%\author[]{Yinhe Han}{}
%\author[]{Qinfen Hao}{}
%\author[]{Shuo Hou}{}
%\author[]{Xuan Huang}{}
%\author[]{Pengwei Jin}{}
%\author[]{Changxin Ke}{}
%\author[]{\\Guangli Li}{}
%\author[]{Huawei Li}{}
%\author[]{Kuan Li}{}
%\author[]{Naipeng Li}{}
%\author[]{Shengwen Liang}{}
%\author[]{Cheng Liu}{}
%\author[]{Hongwei Liu}{}
%\author[]{\\Jiahua Liu}{}
%\author[]{Junliang Lv}{}
%\author[]{Jianan Mu}{}
%\author[]{Jin Qin}{}
%\author[]{Bin Sun}{}
%\author[]{Duo Wang}{}
%\author[]{Chenxi Wang}{}
%\author[]{\\Mingjun Wang}{}
%\author[]{Mingyu Yan}{}
%\author[]{Xiao Xiao}{}
%\author[]{Xiaochun Ye}{}
%\author[]{Chenwei Xiong}{}
%\author[]{Ruiyuan Xu}{}
%\author[]{\\Mengyao Xie}{}
%\author[]{Kuai Yu}{}
%\author[]{Rui Zhang}{}
%\author[]{Shuoming Zhang}{}
%\author[]{Jiacheng Zhao}{}
%\author[2]{Ccc AUTHOR}{}
%\author[3]{Ddd AUTHOR}{}
\author[]{Kaiyan Chang}{}
\author[]{Mingzhi Chen}{}
\author[*]{Yunji Chen}{{cyj@ict.ac.cn}}
\author[]{Zhirong Chen}{}
\author[]{Dongrui Fan}{}
\author[]{\\Junfeng Gong}{}
\author[]{Nan Guo}{}
\author[]{Yinhe Han}{}
\author[]{Qinfen Hao}{}
\author[]{Shuo Hou}{}
\author[]{Xuan Huang}{}
\author[]{Pengwei Jin}{}
\author[]{\\Changxin Ke}{}
\author[]{Cangyuan Li}{}
\author[]{Guangli Li}{}
\author[]{Huawei Li}{}
\author[]{Kuan Li}{}
\author[]{Naipeng Li}{}
\author[]{\\Shengwen Liang}{}
\author[]{Cheng Liu}{}
\author[]{Hongwei Liu}{}
\author[]{Jiahua Liu}{}
\author[]{Junliang Lv}{}
\author[]{Jianan Mu}{}
\author[]{Jin Qin}{}
\author[]{\\Bin Sun}{}
\author[]{Chenxi Wang}{}
\author[]{Duo Wang}{}
\author[]{Mingjun Wang}{}
\author[*]{Ying Wang}{{wangying2009@ict.ac.cn}}
\author[]{Chenggang Wu}{}
\author[]{\\Peiyang Wu}{}
\author[]{Teng Wu}{}
\author[]{Xiao Xiao}{}
\author[]{Mengyao Xie}{}
\author[]{Chenwei Xiong}{}
\author[]{Ruiyuan Xu}{}
\author[]{\\Mingyu Yan}{}
\author[]{Xiaochun Ye}{}
\author[]{Kuai Yu}{}
\author[]{Rui Zhang}{}
\author[]{Shuoming Zhang}{}
\author[]{Jiacheng Zhao}{}
%Kaiyan Chang, Yunji Chen, Mingzhi Chen, Zhirong Chen, Dongrui Fan, Junfeng Gong, Nan Guo, Yinhe Han, Qinfen Hao, Shuo Hou, Xuan Huang, Pengwei Jin, Changxin Ke, Kuan Li, Huawei Li, Guangli Li, Naipeng Li, Cangyuan Li, Shengwen Liang, Cheng Liu, Jiahua Liu, Hongwei Liu, Junliang Lv, Jianan Mu, Jin Qin, Bin Sun, Duo Wang, Ying Wang, Chenxi Wang, Mingjun Wang, Teng Wu, Peiyang Wu, Chenggang Wu, Xiao Xiao, Mengyao Xie, Chenwei Xiong, Ruiyuan Xu, Mingyu Yan, Xiaochun Ye, Kuai Yu, Rui Zhang, Shuoming Zhang, Jiacheng Zhao
%%% Authors' contribution. Í¬µÈ¹±Ï×
\contributions{All authors contributed equally to this work and are listed in alphabetical order.}

%%% Address. µØÖ·
%%%\address[number]{Affiliation, City Postcode, Country}
\address[]{Institute of Computing Technology, Chinese Academy of Sciences, Beijing 100190, China}
%\address[2]{Affiliation, City 000000, Country}
%\address[3]{Affiliation, City 000000, Country}

\input{submit_version/0_abstract}

 \maketitle

\input{submit_version/1_introduction}
\input{submit_version/4_Compiler}
\input{submit_version/5_BinaryTranslation}
\input{submit_version/6_Simulator}
\input{submit_version/7_hw_sw_partition}
\input{submit_version/8_dse}
\input{submit_version/9_rtl}

\input{submit_version/10_sam}
\input{submit_version/11_conclusion}

%%%%%%%%%%%%%%%%%%%%%%%%%%%%%%%%%%%%%%%%%%%%%%%%%%%%%%%
%%% The main text. ÕýÎÄ²¿·Ö
%%%%%%%%%%%%%%%%%%%%%%%%%%%%%%%%%%%%%%%%%%%%%%%%%%%%%%%

\newpage

%%%%%%%%%%%%%%%%%%%%%%%%%%%%%%%%%%%%%%%%%%%%%%%%%%%%%%%
%%% Acknowledgements. ÖÂÐ»
%%%%%%%%%%%%%%%%%%%%%%%%%%%%%%%%%%%%%%%%%%%%%%%%%%%%%%%
%\Acknowledgements{This work was supported by the National Natural Science Foundation of China (Grant Nos. 00000000 and 11111111).}

%%%%%%%%%%%%%%%%%%%%%%%%%%%%%%%%%%%%%%%%%%%%%%%%%%%%%%%
%%% Supplements. ²¹³ä²ÄÁÏ, ·Ç±ØÑ¡
%%%%%%%%%%%%%%%%%%%%%%%%%%%%%%%%%%%%%%%%%%%%%%%%%%%%%%%
%\Supplements{Appendix A.}

%%%%%%%%%%%%%%%%%%%%%%%%%%%%%%%%%%%%%%%%%%%%%%%%%%%%%%%
%%% Reference section. ²Î¿¼ÎÄÏ×
%%% citation in the content using "some words~\cite{1,2}".
%%% ~ is needed to make the reference number is on the same line with the word before it.
%%%%%%%%%%%%%%%%%%%%%%%%%%%%%%%%%%%%%%%%%%%%%%%%%%%%%%%
\bibliographystyle{scis}
\bibliography{main_dse, main}
%\bibliography{main}

%%%%%%%%%%%%%%%%%%%%%%%%%%%%%%%%%%%%%%%%%%%%%%%%%%%%%%%
%%% Appendix sections. ¸½Â¼ÕÂ½Ú, ·Ç±ØÑ¡
%%%%%%%%%%%%%%%%%%%%%%%%%%%%%%%%%%%%%%%%%%%%%%%%%%%%%%%
%\begin{appendix}
%\section{Name}

%\end{appendix}

\end{document}

%% file: submit_version/0_abstract.tex
\abstract{Computer System Architecture serves as a crucial bridge between software applications and the underlying hardware, encompassing components like compilers, CPUs, coprocessors, and RTL designs. Its development, from early mainframes to modern domain-specific architectures, has been driven by rising computational demands and advancements in semiconductor technology. However, traditional paradigms in computer system architecture design are confronting significant challenges, including a reliance on manual expertise, fragmented optimization across software and hardware layers, and high costs associated with exploring expansive design spaces. While automated methods leveraging optimization algorithms and machine learning (ML) have improved efficiency, they remain constrained by a single-stage focus, limited data availability, and a lack of comprehensive human domain knowledge. The emergence of large language models (LLMs) offers transformative opportunities for the design of computer system architecture and search paradigms. By leveraging the capabilities of LLMs in areas such as code generation, data analysis, and performance modeling, the traditional manual design process can be transitioned to a machine-based automated design approach. To harness this potential, we present the Large Processor Chip Model (LPCM), an LLM-driven framework aimed at achieving end-to-end automated computer system architecture design. The development of LPCM is structured into three levels: (1) Human-Centric, which assists in code generation and parameter tuning; (2) Agent-Orchestrated, facilitating cross-layer optimization through toolchain integration (e.g., LLVM, Gem5) and the autonomous execution of subtasks; and (3) Model-Governed, achieving full automation through the synthesis of hardware-software co-design, simulation, and iterative refinement. This paper utilizes 3D Gaussian Splatting (3D GS) as a representative workload and employs the concept of software-hardware collaborative design to examine the implementation of the LPCM at Level 1, demonstrating the effectiveness of the proposed approach. Furthermore, this paper provides an in-depth discussion on the pathway to implementing Level 2 and Level 3 of the LPCM, along with an analysis of the existing challenges.
}

\keywords{Large Processor Chip Model, LLMs, Automated Design, 3D Gaussian Splatting}

%% file: submit_version/1_introduction.tex
\section{Introduction}
%介绍computer system architecture是什么
\textbf{Computer System Architecture} is a fundamental discipline in the realms of computer science and engineering, concentrating on the design, organization, and performance optimization of computer systems. It comprises multiple components including compilers, CPUs, coprocessors, and RTL designs. %Among these, the compiler serves as a bridge between software and hardware, translating high-level languages into the target machine's instruction set architecture (ISA) while performing code optimizations to enhance execution efficiency. The CPU acts as the general-purpose computing core, implementing ISA functionality through microarchitecture features such as pipelining, out-of-order execution, and multi-level caches, while handling logical operations and control flow tasks. Coprocessors are employed to accelerate specific computational workloads, utilizing dedicated hardware circuits to optimize energy efficiency and offload the CPU. Finally, hardware description languages like Verilog are used to define the logical behavior of digital circuits, encompassing datapaths, control units, and sequential logic, which are ultimately synthesized into the physical layout of actual chips. Computer System Architecture serves a pivotal bridging role in connecting upper-layer applications with the underlying hardware technologies. 
From a software perspective, computer system architecture follows the layered architecture design philosophy and establishes multi-layer abstraction to bridge the gap between upper-layer applications and underlying physical hardware, providing a stable development environment for applications. From a hardware perspective, the evolution of computational demands in contemporary applications, coupled with advancements in algorithmic, continuously challenge the performance boundaries of underlying hardware, driving sustained innovation in underlying hardware technologies. As a bridge between software applications and hardware technologies, computer system architecture fundamentally influences the performance, energy efficiency, reliability, and scalability of computing systems. The evolution and innovation in computer system architecture have become key drivers in advancing computing technology, playing a essential role in fostering technological innovation and addressing diverse computational demands.

%介绍computer system architecture 研究的发展
\subsection{The Evolution of Design Paradigms in Computer System Architecture}

Since the advent of the first electronic computer ENIAC, in the 1940s, computer system architecture has undergone a structural evolution from a single form to a highly diversified landscape. From the 1940s to the 1960s, mainframe computers represented by ENIAC dominated the field of computing. %They were scarce in number, extremely expensive, and used exclusively for military and scientific research purposes. Their system architecture primarily consisted of arithmetic circuits with a simple, monolithic structure. Programs were input via punched paper tapes, and the concept of software had not yet emerged.
With the advent of microprocessors and breakthroughs in integrated circuit technology, the 1970s to 1990s saw the widespread adoption of microcomputers and personal computers. The physical size of computing devices shrank to about 0.25 square meters, costs dropped to several hundred US dollars, and computing performance reached hundreds of millions of operations per second. %The types of devices diversified from single mainframe configurations to multiple forms including PCs and servers. The number of computing devices grew exponentially, e.g. global PC sales were just a few million units in 1983 but exceeded 100 million units by 1999. 
During this period, computer architectures became increasingly complex due to functional generalization and performance demands, making hierarchical design and hardware-software co-optimization mainstream. To improve development efficiency, high-level languages such as C and compiler technologies advanced rapidly, freeing programmers from the burden of direct hardware manipulation. On the hardware side, considerations extended beyond arithmetic units to include CPU microarchitecture design elements like pipelines, multi-level caches, I/O systems, and bus architectures.

Entering the 21st century, especially after the resurgence of deep learning algorithms in 2012, %with the explosive growth of mobile computing and the Internet of Things (IoT), computing device forms expanded from traditional PCs to various specialized domains including smartphones, wearables, and smart sensors. Since 2014, the number of deep learning processor chips in the world has grown from one or two to hundreds of types \cite{Guo2024Accelerator}, covering  various specialized categories from industrial sensors to wearables and smart home devices. In this era, general-purpose CPUs could no longer meet the demands of diverse application scenarios, making application-specific coprocessors the mainstream design approach.
the rapid development of emerging applications and advancements in semiconductor technology have shifted the focus from general-purpose architectures to domain-specific computer system \cite{hennessy2019new}. Despite Moore's Law witnessed the reductions of the chip manufacturing cost, the escalating demand for specialized hardware tailored to a wide array of application requirements such as energy efficiency constraints, real-time performance requirements, has led to a substantial rise in chip development costs. However, current computer system architecture design not only heavily relies on domain experts but also suffers from long design cycles, making it difficult to meet the demands of rapidly evolving applications. Moreover, computer system architecture design encompasses a comprehensive process from high-level software interfaces to low-level hardware implementations, including compiler design, operating system optimization, hardware-software partitioning, micro-architecture design, RTL design, simulation, and verification, among other critical stages. This results in a vast design space with multi-objective optimization challenges, where traditional reliance on manual expertise severely constrains design performance and outcomes.

%系统结构自动化设计的重要性
Automated computer system architecture design technology is an effective approach to enhancing the performance and efficiency of system architecture design. Automated design leverages various optimization algorithms, particularly those rooted in artificial intelligence, to rapidly explore the design space and achieve hardware-software co-optimization. This approach improves design performance, shortens design cycles, reduces development costs, and better meets the demand for customized system architectures in emerging fields such as artificial intelligence and high-performance computing. As a result, automated computer system architecture design is not only a key solution to current design challenges but also a significant driver of future computing technology innovation. Throughout the evolution of computer system architecture, the field has transitioned from manual design to automated design, with the capabilities of automation continuously strengthened by advancements in machine learning technologies, particularly deep learning. The evolution of automated design can be divided into three main categroies.

%回顾发展历程：从手工设计到自动设计，ml设计，自动化程度越来越高，自动化的分级
%TODO：这部分的问题：
%1、主要内容是处理器设计，还需要补充编译器和OS方面内容，应该需要专门写一章related work，分开编译器、OS、处理器分别介绍
%2、阶段划分是否合理，是否会有重叠
\paragraph{Traditional Chip Logic Design Based on EDA (Electronic Design Automation) Tools}
Traditional chip logic design based on EDA tools marks the starting point of automated design, signifying the transition from purely manual design to automated processes. Designers decompose and design internal functional modules of computer system architectures according to requirement specifications, translate these designs into hardware description languages (HDLs), and then utilize EDA tools for synthesis, verification, and analysis to generate logic circuits and ultimately physical layouts \cite{ABC} \cite{EdwardsOpenCircuit} \cite{OpenSTA} \cite{IcarusVerilog} \cite{Yosys}. The introduction of EDA tools has enabled complex chip design tasks to be carried out more efficiently and accurately, significantly advancing the development of integrated circuits and computer system architectures. While traditional chip logic design based on EDA tools laid the foundation for modern automated computer system architecture design, this stage of design remained heavily reliant on manual rules and expertise, with the level of automation being highly limited.

\paragraph{Optimization-Based Methods}
Optimization-based methods represent a shift from traditional EDA-assisted automation to more intelligent and refined design optimization processes. At this stage, designers introduce advanced mathematical optimization techniques to explore a broader design space for optimal solutions while ensuring design accuracy and manufacturability. For example, design space exploration (DSE) employs heuristic search methods \cite{schafer2009adaptive} \cite{mahapatra2014machine} to identify the best trade-offs among performance, power, and area during architectural design and microarchitecture optimization. Logic synthesis methods generate optimized logic circuits through Boolean optimization \cite{brayton1984logic}, technology mapping \cite{keutzer1987dagon}, and Bayesian optimisation \cite{grosnit2022boils}. In the physical design phase, methods like placement and routing \cite{cheng2018replace} and timing optimization \cite{singh1988timing} are used to optimize the physical implementation of processors \cite{ajayi2019toward}. These advancements have not only significantly improved design efficiency and precision but also laid the groundwork for subsequent machine learning-based design methods. However, these optimization-based approaches still rely heavily on human expertise, exhibit limited capabilities in handling complex designs, and often require substantial computational resources and optimization time, which can impact the effectiveness of the optimization process.

\paragraph{Machine Learning-Based Methods}
Machine learning-based methods represent a paradigm shift in automated design, transitioning from rule-driven and optimization-driven approaches to intelligent, data-driven methodologies. By learning from historical design data, machine learning can automatically generate new design solutions and even predict and address issues that are challenging for traditional optimization methods. For instance, techniques such as random forests \cite{carrion2012machine} \cite{zuluaga2012smart} and neural networks \cite{ferianc2020improving} have been integrated into design space exploration to predict the performance, power, and area of different configurations, thereby accelerating the exploration process and more efficiently identifying near-globally optimal solutions. Furthermore, reinforcement learning and neural networks (NNs) have been applied to physical design.  In placement, reinforcement learning algorithms iteratively discover optimal micro placement strategies to minimize signal delay and congestion \cite{mirhoseini2020chip} \cite{he2020circuit}. NNs, on the other hand, contribute to performance prediction of timing, power, and congestion \cite{alawieh2020high}.
While these techniques apply machine learning methods to boost EDA efficiency or performance, they do not fundamentally transform the conventional design flow.

%3.自动设计的难点和挑战：（待定）
%(1)多个层级阶段，互相影响和制约，需要综合考虑（为跨层优化做铺垫）
%(2)难以利用人类知识，涉及到的领域知识较多且庞杂（为使用大模型做铺垫）
%(3)数据稀缺
%\subsection{Challenges}
\subsection{The key challenges preventing the evolution of design paradigms}
%横向扩展（晶体管规模，模块复杂度）纵向发展（软硬件栈越来越深，app,os,lib,isa等等）导致的设计成本人月，钱上升趋势，以及更加难以协同（design loop的变长）等等，并以nvidia orion汽车芯片开放成本为例，

Despite the significant advancements brought by machine learning methods to the automated design of computer system architectures, several challenges and issues remain.

First, computer system architecture encompasses a comprehensive design flow from high-level software interfaces to low-level hardware implementation, including stages such as compiler design, operating system support, processor architecture design, circuit design, physical design, verification, and evaluation. These stages are interconnected and mutually constrained, requiring a holistic consideration of their characteristics to achieve globally optimal designs. 
%However, existing automated design methods are often limited to optimizing specific design steps, making it difficult to perform end-to-end optimization across multiple stages. This limitation restricts the potential for optimizing overall system performance.
Nevertheless, existing automated design approaches are typically limited to single-stage optimization and cannot simultaneously handle multiple design stages, making end-to-end cross-stage optimization unattainable, and consequently restricting overall system performance improvement potential.

Second, traditional human-driven design processes have accumulated a wealth of design experience and rules, which provide valuable insights for achieving automation. However, current mainstream machine learning methods primarily rely on automatically extracting patterns from data, failing to effectively incorporate human design expertise and rules. Moreover, computer system architecture design involves multiple stages, each with its own unique design experiences and rules. The complexity and diversity of domain knowledge pose significant challenges for integrating human design knowledge into automated methods.

%Finally, machine learning-based methods depend on large volumes of high-quality data for model training, and the quantity and quality of data directly determine the effectiveness of the models. However, computer system architecture design is characterized by its high specialization and complexity, with relevant data typically generated only by domain experts. This results in limited data availability and high collection costs, further constraining the development and application of machine learning-based automated design methods.

%4.大模型背景-大模型为自动设计带来新的希望
%大模型发展历程，copilot方法，agent，领域专用llm等
\subsection{Large language models bring new opportunities}
Recently, the emergence of large language models (LLMs) and agent systems has prompted researchers to explore the potential for automating the design of computer system architectures. These methods focus on utilizing LLMs to transform natural language descriptions of functional requirements or documentation into appropriate computer system architecture designs, which can lower the barriers to hardware development and improve the efficiency of research and development efforts. 

Research focusing on LLM-driven technical solutions can be categorized into several key areas. First, studies targeting the generation of processor components concentrate on designing various functional modules of processors. Due to the scarcity of data in the processor design domain, these works propose diverse methods for constructing processor design datasets such as ChipNeMo \cite{ChipNeMo24}, RTLCoder \cite{RTLCoder23} and fine-tune large language models such as BetterV \cite{BetterV24} on these datasets to obtain specialized hardware code generation models capable of producing Verilog or VHDL code.
Second, research aimed at designing comprehensive frameworks focuses on building system-level automated design workflows. These works implement complex or system-level hardware architecture design such as ChatEDA \cite{wu2024chateda} and ChipGPT \cite{ChipGPT23}, through modules such as task decomposition and composition, code verification and feedback, performance optimization, and search strategies. Additionally, these methods often incorporate automatic feedback and correction mechanisms, using simulation or formal verification tools to detect errors in the generated code and iteratively optimize code quality.
Finally, research targeting analysis and verification tools utilizes large language models to understand, analyze, and summarize hardware code, serving as verification tools to automatically identify complex errors and security issues. Building on this, methods for automatically repairing erroneous code or optimizing power, performance, and area (PPA) have been proposed, e.g., AssertLLM \cite{fang2024assertllm} and RTLFixer \cite{tsai2024rtlfixer}. 
These advancements demonstrate the potential of LLMs in automating and enhancing various aspects of hardware design, from component generation to system-level frameworks and verification tools.

%现有工作的局限性
%(1)依然以单个阶段为主，未进行多个层级的整体设计
%(2)正确性难以保证
Despite the significant potential demonstrated by large language models (LLMs) in computer system architecture  automated design, existing approaches still suffer from several critical limitations. Firstly, current methods primarily focus on individual design stages, such as RTL-level hardware code generation or code verification, and fail to achieve a multi-level, holistic design process that spans from high-level requirements to low-level implementation. Moreover, they lack the capability to enable co-design across software and hardware layers, including compilers, operating systems, and processor design. This limitation results in a lack of end-to-end co-optimization in the design flow, making it challenging to achieve global optimization across performance, power, and area (PPA) objectives. Secondly, the correctness of the design outcomes remains difficult to guarantee. Although LLMs can generate syntactically correct hardware code, the complexity of hardware design and the stringent timing and resource constraints often lead to logical errors or functional defects in the generated code. As a result, manual verification or formal tools are still required to correct these issues. These shortcomings restrict the practical application of existing LLM-based automated design methods in real-world engineering scenarios.

\subsection{Towards next-generation paradigm: Large Processor Chip Model}
%缺一个图，就是整个系统system and arch设计flow的图，从后面挑一个前置，同时，几个阶段分类定义的表也前置放到这里，后面的分级定义有定重复，也有不一致，要改

%5.系统结构大模型的可行性和优势：（待定）
%(1)学习大量知识的能力，
%(2)涌现能力，可以通过贯穿多个层次，最终实现超越人类设计
To harness the powerful capabilities of large language models (LLMs) for the automated design of computer system architectures, we propose the Large Processor Chip Model (LPCM), which is built on LLM technology and domain-specific data from computer system architecture. LPCM aims to achieve end-to-end automated design of computer system architectures. The design of LPCM offers unique advantages and holds significant importance for realizing automated computer system architecture design.
First, leveraging the robust knowledge-learning capabilities of LLMs, LPCM can extract and utilize human expertise and knowledge in system architecture design from vast datasets, such as rules and heuristics in compiler optimization, microarchitecture design, and physical implementation. This capability enables LPCM to quickly grasp complex design logic and apply it to automated design workflows. Second, when the model's parameter size and data volume reach a certain scale, LLMs exhibit emergent abilities, leading to qualitative leaps in performance and behavior. Building on the emergent capabilities of LLMs, LPCM can perform multi-level, cross-domain co-optimization by simultaneously considering design constraints and objectives across multiple layers, such as compilers, operating systems, hardware architectures, and physical implementation. This enables LPCM to generate system architecture designs that surpass human capabilities. Such global design capabilities not only significantly improve design efficiency but also achieve better trade-offs among performance, power, and area (PPA) objectives, driving computer system architecture design toward higher levels of intelligence and automation.

%7.sam的整个flow和功能，
To realize LPCM, multiple modules need to be collaboratively designed to cover the full technology stack from high-level software interfaces to low-level hardware implementation. This includes several critical components such as compiler design, operating system support, hardware-software partitioning, microarchitecture design, RTL (Register Transfer Level) design, and simulator development. 
The compiler translates high-level language programs into machine instructions, enabling software to interact efficiently with hardware. The operating system manages hardware resources and provides abstract interfaces, ensuring efficient resource allocation and system stability. Hardware-software partitioning determines whether specific functionalities are implemented in hardware or software, balancing performance, flexibility, and design complexity. Microarchitecture design focuses on optimizing processor performance and energy efficiency by defining the internal structure and data flow of the processor. RTL design implements the hardware logic, specifying the behavior of digital circuits at the register transfer level. Finally, simulators are used to verify and evaluate system performance, ensuring that the design meets functional and performance requirements before physical implementation.
By integrating these modules, LPCM can achieve a comprehensive and automated design flow, enabling end-to-end optimization and validation of computer system architectures. This holistic approach not only enhances design efficiency but also ensures that the final system meets the desired performance, power, and area (PPA) objectives.

Achieving LPCM is a highly challenging task, whose complexity and interdisciplinary nature determine that this goal cannot be accomplished overnight but requires gradual advancement in stages. Based on the varying degrees of automation in the design process, the implementation can be divided into the following three levels.

\textbf{Level 1: Human-Centric Hierarchical Design and Optimization.} In the human-centric hierarchical design and optimization phase, LPCM serves as auxiliary tools to assist humans in designing computer system architectures. Humans, as the primary drivers of the design process, are responsible for setting goals of hierarchical design, such as instruction set architecture, memory hierarchy, compiler optimization strategies, and more. They accomplish the main design tasks with the help of existing system architecture design tools like LLVM, GEM5, Chisel, and Verilog simulators. The primary role of LPCM in this phase is to provide suggestions to humans, such as generating code snippets, proposing optimization algorithms, or offering hardware description language (HDL) templates. However, for the design of complex components, such as processor pipelines and cache coherence protocols, the decision and optimization still heavily rely on human expertise.

During this level, human involvement accounts for a significant portion of the work, and the use of design tools is frequent. The contributions of LPCM are limited, primarily focusing on repetitive tasks like code generation and parameter tuning, or knowledge retrieval tasks such as searching for relevant research papers, tool documentation, or design specifications. Human experts are responsible for the hierarchical verification and optimization of the design results, ensuring that the design goals at each level, such as compiler layer, hardware architecture layer, and hardware module layer, are consistent and efficient.

In this level, LPCM possess limited domain-specific knowledge. It may directly employ general-purpose large language models like ChatGPT or DeepSeek without undergoing deep fine-tuning for the computer system architecture domain or integration with domain-specific tools like LLVM, Gem5, or Chisel. As a result, the outputs of these models require rigorous review and optimization by human experts, especially in cross-layer designs such as hardware-software co-design.

\textbf{Level 2: Agent-Orchestrated Cross-Layer Design and Optimization.} In the agent-orchestrated cross-layer design and optimization phase, LPCM act as intelligent agents capable of independently completing certain subtasks, such as automatically generating compiler optimization passes, designing simple processor microarchitectures, and creating operating system scheduling algorithms. Humans only need to define the design objectives for these subtasks, such as performance metrics and power constraints, while LPCM autonomously handle the design, evaluation, and error correction of these subtasks. For example, through domain-specific fine-tuning, LPCM can generate hardware description language (HDL) code, design pipeline configurations, and cache hierarchies for processor microarchitectures. For cross-layer optimization tasks like hardware-software co-design, LPCM can coordinate design goals across different layers, such as the compiler layer, hardware architecture layer, and hardware module layer, and automatically integrate and invoke toolchains like LLVM and Gem5 to accomplish specific tasks. In terms of evaluation, LPCM can utilize simulation tools such as Gem5 and Verilog simulators to assess hardware design performance, generate performance reports on throughput, latency, and power consumption, and conduct corresponding analyses. LPCM can also automatically detect issues in the design, such as performance bottlenecks, functional errors, or power consumption violations, and generate corrective solutions, such as adjusting pipeline configurations, modifying cache coherence protocols, or optimizing scheduling algorithms, followed by regenerating the design. However, for highly complex tasks, such as designing multi-core cache coherence protocols, human expert intervention and optimization are still required.

During this level, human involvement significantly decreases, and the reliance on design tools is reduced, while the responsibilities of LPCM increase substantially. LPCM play a central role in exploring the design space and can rapidly generate and evaluate multiple design solutions, such as different processor pipeline configurations and memory hierarchies. Additionally, LPCM can automate verification tasks, including formal verification and simulation testing, to ensure design consistency and correctness.

In this level, LPCM possess extensive domain-specific knowledge, having been fine-tuned using domain data such as research papers, open-source projects, and toolchain documentation from the computer architecture field. They are capable of understanding and generating code and design documents that comply with domain-specific standards. Furthermore, LPCM can propose optimization suggestions based on cross-layer design requirements, such as matching compiler optimizations with hardware characteristics or coordinating operating system scheduling with processor microarchitecture, thereby achieving overall performance improvements.

\textbf{Level 3: Model-Governed Autonomous Design and Optimization.} In the model-governed autonomous design and optimization phase, LPCM achieve full automation of the entire design process, independently completing all subtasks such as compiler optimization, operating system scheduling, instruction set architecture design, and processor microarchitecture design. LPCM also establish a complete closed loop encompassing system design, evaluation, and error correction. LPCM can autonomously invoke toolchains like LLVM, Gem5, and Verilog simulators to perform design, simulation, verification, and optimization, entirely replacing manual design and the use of traditional design tools. Through cross-layer optimization in hardware-software co-design, LPCM ensure consistency in design goals across all levels and achieve optimal overall performance.

During this level, humans only need to propose high-level objectives, such as ``design a power-efficient RISC-V processor’’, and LPCM can autonomously complete the entire design process. The involvement of LPCM approaches 100\%, with almost no need for human intervention. LPCM can automatically decompose tasks, generate design solutions, and iteratively optimize them using simulation and verification tools like Gem5 and Synopsys VCS. Based on design goals and constraints such as performance, power consumption, and area, LPCM can explore the design space and identify optimal solutions.

In this level, LPCM possess extensive domain-specific knowledge, reaching or even surpassing the level of domain experts in terms of design efficiency and quality. It can integrate the latest research advancements, such as novel memory technologies and quantum computing architectures, to propose innovative design solutions, driving the forefront of computer system architecture development.

%Please use this sample as a guide for preparing your letter. Please read all of the following manuscript preparation instructions carefully and in their entirety. The manuscript must be in good scientific American English; this is the author's responsibility. All files will be submitted through our online electronic submission system at \href{https://mc03.manuscriptcentral.com/scis}{HERE}.

%% file: submit_version/4_Compiler.tex
\section{Compiler meets LLM}

\subsection{Motivation}
\label{sec:compiler_overview}
Compilers play a crucial role as crucial intermediaries between human-readable source code and machine-executable code, ensuring reliable application performance across diverse hardware architectures. 
With the emergence of domain-specific applications, specialized architectures have been designed to enhance performance and reduce power consumption. 
However, significant challenges persist in the collaborative design of software and hardware, as iterative updates typically demand months or even years to complete.  
This bottleneck has prompted researchers to investigate automated methods for compiler construction.

The traditional compiler toolchain process, ranging from lexical analysis and parsing to optimization and code generation, is well-established but demands substantial expert knowledge, especially when adapting to new architectures.
This dependency results in a bottleneck, primarily attributed to the scarcity of experienced compiler engineers.
The LLM compiler is designed to tackle this limitation by automating the adaptation of compilers for emerging domain-specific architectures, thereby facilitating rapid development iteration cycles between hardware modules and software modules within the LPCM framework.

\begin{figure}
\includegraphics[width=0.98\textwidth]{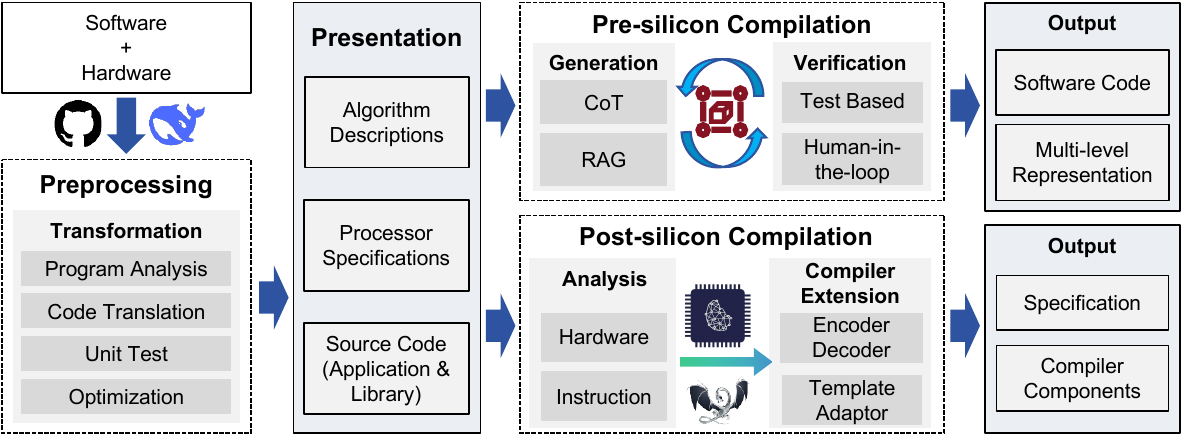}
\caption{LLM Compiler in Large processor chip model: an overview}
\label{fig:compiler:overview}
\end{figure}

Large language models have demonstrated exceptional capabilities in code generation and pattern recognition, offering a promising approach to addressing the challenges in compiler development.
Within the LPCM framework, the compiler module serves as a critical bridge connecting software applications, hardware specifications, and hardware architectures.
This connection has the potential to significantly accelerate development cycles by enabling automated transformation and optimization. 
While AI-powered tools like GitHub Copilot, TabNine, and Cursor have enhanced general code development productivity, they often exhibit limitations in comprehending complex system architectures and compiler design principles.
However, their effectiveness remains constrained by limited exposure to compiler source code during training.
Thus this field continues to depend heavily on scarce expert knowledge.
The LLM compiler approach aims to address this limitation by developing specialized models with deeper understanding of compiler construction.
This not only alleviates the engineering bottleneck but also enables more efficient code generation tailored for iteratively evolving hardware architectures.

There is an increasing demand for specialized large language models in compiler-related tasks within system architecture research.
These expert-level LLMs are crucial for adapting to changes in emerging domain-specific architectures, facilitating collaboration between hardware and software development, and enhancing the co-design process within the LPCM.
By training with datasets that incorporate compiler knowledge, such as source codes and assembly instructions, the automation and intelligence of system software within a LPCM can be significantly enhanced, thereby improving development workflows significantly.
When a new ISA is produced—either a novel design or an extension—there is typically no immediately available compiler toolchain to support it.
To this end, the compiler module must undergo iterative updates to ensure efficient code generation and execution tailored to the new architecture.
The connection pathway of the compiler module with its adjacent components within the LPCM framework is described as follows:
\textit{1) Inputs.} The compiler module receives the software source code from upstream modules containing new algorithm implementations and partition information, and hardware basic specifications for further implementation, which include parameters that the compiler should consider during the generation and optimization process.
\textit{2) Outputs.} The compiler module generates the transformed source code and appropriate intermediate representations for downstream modules, such as DSE modules, and constructs compiler components that can be integrated into conventional compiler systems during iterative design refinement.

\subsection{The overview of LLM Compiler}
To leverage the capabilities of LLMs and realize fully automated compiler design without human intervention, we propose the \textbf{LLM Compiler}, an advanced compilation toolchain powered by LLMs, tailored specifically for compilation and programming tasks within LPCM.
However, achieving the goal of unmanned intervention is a long-term process that cannot be accomplished overnight. 
To support this objective, we define three levels based on technological development trends and the varying capabilities of LLMs as follows:
\begin{itemize}
    \item \textbf{Level 1: Assisted Compiler Development}. In this process, LLMs utilize compilation knowledge obtained during training or fine-tuning, take engineers' requirements and the code context to be edited as input, and produce edited code as output, including source code analysis and transformation, as well as suggestions for compiler component generation. This capability effectively accelerates engineers' development work, as demonstrated by systems like Cursor.
    \item \textbf{Level 2: Semi-Autonomous Compiler Components construction}. LLMs can fulfill one or more critical roles in three primary compiler tasks: decision-making, integration with the existing compilation system software, and execution of specialized compiler functions. By employing an agent-based framework, the entire workflow, spanning from application intake to compilation and hardware adaptation, can be partially automated. However, human intervention remains essential for refining and adjusting inputs and outputs across modules, as the compilation process is inherently error-prone.
    \item \textbf{Level 3: End-to-End Compiler Generation and Execution}. LLMs can autonomously manage the entire compilation and deployment process. Human users are only required to specify task objectives or provide inputs, after which the system delivers the final results.
\end{itemize}

\textbf{Related Work.} 
Most contemporary research and development efforts remain at Level 1, where LLMs provide supportive functions while humans maintain primary control over the compiler development process.
Representative code completion and automatic programming tools, such as GitHub Copilot and TabNine, belong to this category. 
These tools generate code snippets and provide completion suggestions but exhibit limited understanding of the knowledge in compiler domains.
The source code translation approaches presented in \cite{smt-ASE15} and \cite{tree2tree-NIPS18}, as well as code LLM models such as CodeBERT \cite{feng2020codebert} and CodeT5 \cite{wang2021codet5}, similarly function at Level 1. 
These methods provide support to developers but lack autonomous decision-making capabilities. 
Emerging work is progressively advancing to Level 2, where LLMs assume semi-autonomous roles in constructing compiler components.  
ComBack \cite{zhong2024comback} showcases the LLM-driven automation of instruction selection and register allocation, whereas specialized models introduced in \cite{zhang-etal-2024-introducing} are capable of translating high-level code into extended ISA instructions with minimal supervision.
VeGen \cite{chen2021vegen}, which focus on SIMD vectorization, also belong to this category. These approaches partially automate compiler components but still require human intervention for seamless integration.
Systems at Level 3, which are capable of end-to-end compiler generation and execution with minimal human intervention, remain largely theoretical. 
Recent neural compilation methods \cite{armengol2021learning, c2llvm-HPEC22} provide promising foundations for potentially realizing such systems.
The progression from Level 2 to Level 3 signifies the current research frontier, demanding substantial advancements in LLMs' capabilities to reason about intricate compiler architectures, manage cross-component interactions effectively, and produce both reliable and optimized code generation pathways.

Our LLM Compiler, currently at Level 2 and targeting advancement to Level 3, encompasses two distinct approaches specifically designed to meet different development requirements: \textbf{LLM as Compiler} and \textbf{LLM generates Compiler}.

\subsubsection{LLM as Compiler}
\label{sec:compiler:llm_as_compiler}

LLMs can directly serve as translation components for converting source code to an extended ISA design, functioning as an LLM Compiler. In this design paradigm, the process differs based on two distinct scenarios. In the first scenario, when working with an existing Domain-Specific Architecture (DSA) that incorporates new hardware components, the LLM Compiler identifies code segments suitable for acceleration on the new hardware while maintaining compatibility with the established DSA framework. In the second scenario, when confronting an entirely new DSA or significant instruction set extensions, the LLM Compiler must perform comprehensive analysis and translation of the source code to fully leverage the novel architectural capabilities, effectively bridging the gap between conventional programming paradigms and the innovative instruction set.
% LLMs can directly serve as translation components for converting source code to an extended ISA design, functioning as an LLM Compiler. In this design paradigm, the process begins with the identification and analysis of the source code, followed by the direct translation of the identified segments of the source code tailored for the new ISA.
% This approach is particularly suitable for domain-specific architecture design, as it does not require compatibility with general-purpose architectures. Below, we will outline our specific vision for this approach.
% 现有dsa但是新硬件/新dsa（扩展指令）

The LLM first functions as an analyzer, assessing the characteristics of the source code and partitioning it into multiple sub-regions (using either basic blocks, control blocks, or functions as granular units). For each sub-region, the LLM evaluates whether it can be mapped to extended instructions or operators. If mapping is feasible, we assume that translating to extended instructions will yield benefits. The cost model provided by the ISA design is then used to select the mapping approach that offers the greatest advantage.

Subsequently, the LLM acts as a translator, translating the source code according to the selected mapping scheme. It is important to recognize that this translation process is prone to errors; therefore, it requires constraint information and examples from the ISA design as few-shot guidance. Additionally, the results must be verified for functional equivalence against the original program—specifically, by comparing the output of the compiled code to that produced when executed on a pure CPU—to ensure the correctness of the generated code.

Looking toward the future, LLM Compilers within LPCM will likely evolve to incorporate reinforcement learning from verification results, allowing them to continuously improve translation accuracy and optimization strategies. They may also develop capabilities to suggest architectural modifications based on observed software patterns, effectively participating in the co-design process rather than merely implementing it. As these models mature, they could potentially generate entire compiler toolchains for new architectures automatically, eliminating one of the major bottlenecks in deploying novel computational paradigms.

\subsubsection{LLM generates Compiler}
\label{sec:compiler:llm_generates_compiler}

LLMs can also function as code generators, contributing to the development of compiler components.
Traditionally, as hardware architectures evolve through successive iterations, adapting compiler backends to new ISAs requires manual modifications, leading to substantial development costs. 
However, we observe that despite varing ISA features, compiler backends  exhibit consistent structural patterns and invariant operational norms across their analysis and transformation passes.This consistency provides the foundation for LLM-automated compiler generation.

Using LLVM as an example, its compiler backend processes maintain four canonical processing phases regardless of target architecture: instruction selection, register allocation, instruction scheduling, and code emission. Each phase follows deterministic implementation patterns while utilizing Architecture-specific information and stable transformation algorithms. The hardware characteristics required by these phases are uniformly described in target description \texttt{.td} files using TableGen, a DSL. 

However, LLM faces challenges when generating system-level compiler code, especially when processing natural language requirements that often fail to highlight extended instruction set specifications. 
To address this challenge, we propose a systematic approach of analyzing existing LLVM-supported ISAs to construct structured few-shot examples linking natural language requirements, TableGen modifications, and Pass adaptation code generation. 
This framework aims to effectively guide LLMs in extending the LLVM compiler for new ISAs.

Furthermore, the characteristics of compiler backends allow us to extract specific implementation patterns from the workflow in LLVM. We aim to build a compiler backend-oriented fine-tuning dataset to enhance open-source LLM's accuracy in generating compiler backend code.

In future work, these capabilities will be integrated into LPCM, supporting automated adaptation of the compiler module, thereby enabling prototype architecture to be rapidly assessed.  
This approach complements the LLM as Compiler method by addressing distinct phases of the compiler development lifecycle within the LPCM, facilitating rapid hardware-software co-evolution.

% \begin{figure}
% \includegraphics[width=0.98\textwidth]{figure/compiler_future_dataset.pdf}
% \caption{Proposed LLM Compiler dataset construction method}
% \label{fig:compiler:dataset}
% \end{figure}

%% file: submit_version/5_BinaryTranslation.tex
\section{Binary Translation meets LLM}

\subsection{Motivation}

The rapid evolution of processor architectures has brought unprecedented challenges and opportunities to software ecosystem construction for emerging processors. The binary translation module is designed to enable seamless migration and efficient execution of applications across heterogeneous Instruction Set Architectures (ISAs), thereby breaking down ecosystem barriers and accelerating the adoption of new processor platforms. 

Traditional binary translation approaches face two main challenges: (1) \textbf{High development cost and long cycles}, as translators must be manually crafted and tuned for every new ISA combination and scenario, and (2) \textbf{Limited adaptability and scalability}, with heavy reliance on expert knowledge and hand-crafted rules, making it hard to keep up with the rapid proliferation of ISAs and emerging hardware-software co-design requirements. These challenges underscore the urgent need for automated, intelligent binary translation that can efficiently adapt to new architectures and workloads.

What sets our module apart is its integration of Large Language Models (LLMs) to achieve automation throughout the entire binary translation workflow. Leveraging LLMs, the module can not only automate the generation of binary translators but also deliver tailored outputs—such as customized instruction sets and hardware optimization suggestions—based on dynamic program behaviors and hardware features. This marks a paradigm shift from traditional, labor-intensive approaches to a data-driven, intelligent automation framework.

Within the LPCM framework, our module plays a pivotal role: it aims to dramatically reduce development time, automate the generation and optimization of binary translators, and provide the critical outputs as shown in Figure~\ref{fig:4_2_flow_of_bin}. which include key  information—including binary translation hardware extension descriptions, hardware-software co-optimization suggestions, and tailored instruction sets. This information  can be delivered to the modules of Architecture simulator and Design Space Exploration for simulation and implementation, supporting upstream hardware design optimization and downstream software migration and deployment, and facilitating efficient interaction among LPCM modules. 

\subsection{Overview of binary translation tool}

\subsubsection{Development Stages and Related Work}
Across the three levels of LPCM development, we systematically leverage LLMs to introduce intelligence into the binary translation module, focusing on the unique needs and characteristics of binary lifting, instruction mapping and equivalence transformation, as well as dynamic behavior analysis and hotspot detection.

\begin{itemize}
    \item \textbf{Level 1: LLM-Assisted Binary Analysis and Lifting}. In this level, the binary translation toolchain is mainly developed and refined by human experts, with LLMs serving as auxiliary analysis engines. The LLM assists with tasks such as binary lifting—converting binaries to intermediate representations (e.g., LLVM IR)—and the extraction of reduced instruction sets for application-specific translation. The output includes a manually crafted binary translator and a reduced instruction set, which is provided to the Design Space Exploration (DSE) module. 
    
    \item \textbf{Level 2: Agent-Orchestrated Binary Translation and Optimization}. At this level, LLMs evolve from auxiliary tools to orchestrating agents, coordinating multiple submodules such as program analysis and hardware optimization. The LLM-driven agent invokes relevant analysis tools, identifies hot code regions, performs instruction slicing, and generates hardware optimization descriptions. 
    
    \item \textbf{Level 3: End-to-End Autonomous Binary Translator Generation and Optimization}. In this advanced level, the LLM autonomously governs the entire pipeline, model-governed autonomous design and Optimization from binary analysis to translator generation, based on architectural models of both source (guest) and target (host) processors. This enables the rapid, automatic production of highly compatible translators and tailored hardware/software optimization recommendations. Although few current systems have reached this level, it represents the ultimate goal of fully automated, intelligent binary translation.
\end{itemize}

\textbf{Related Work.}  Situating recent binary translation research within our three-level automation framework helps clarify the state of the field and the trajectory of progress. The majority of contemporary efforts remain at level 1, where the core binary lifting and translation toolchains are still primarily developed by human experts, with LLMs or AI serving as auxiliary engines. Tools and frameworks such as BOLT \cite{bolt2023} and Lightning \cite{panchenko2021lightning} exemplify this phase: they leverage sampled runtime information and predefined rules for code layout optimization or binary rewriting, but the pipeline is fundamentally rule-based, requiring extensive manual engineering to adapt to new architectures or applications. Recent years have witnessed a growing body of research advancing towards level 2, where LLMs and machine learning models orchestrate and automate increasingly significant portions of the binary translation process. For example, Wong et al. \cite{wong2023refining} employ GPT-4 as the core of an end-to-end decompilation framework, enabling self-refinement of generated code. Projects like Forklift \cite{armengol2024forklift} and LLM4Decompile \cite{tan2024llm4decompile} illustrate alternative strategies: Forklift trains an end-to-end code-lifting model with enhanced scalability across architectures, while LLM4Decompile augments end-to-end models with refinement modules built atop traditional tools, demonstrating that targeted refinement can outperform naive end-to-end approaches. True level 3 systems—where models autonomously govern the entire binary translation and optimization pipeline, from analysis to code generation and hardware adaptation—are still largely aspirational. 

\subsubsection{Proposed framework design}

Our module aims to drive the transition towards LLM-governed binary translation systems within LPCM. We plan to:

\begin{itemize}
    \item Develop a unified LLM-driven framework that adapts to diverse ISAs and application domains, supporting both static and dynamic translation needs.
    \item Deeply integrate program behavior analysis and hardware co-design, supplying actionable insights for both software migration and hardware optimization.
    \item Continuously provide high-quality, standardized outputs—customized translators, optimization suggestions, and tailored ISAs—to other LPCM modules, expediting system-wide co-evolution and ecosystem expansion.
\end{itemize}

\begin{figure}[htbp]
    \centering
    \includegraphics[width=1\linewidth]{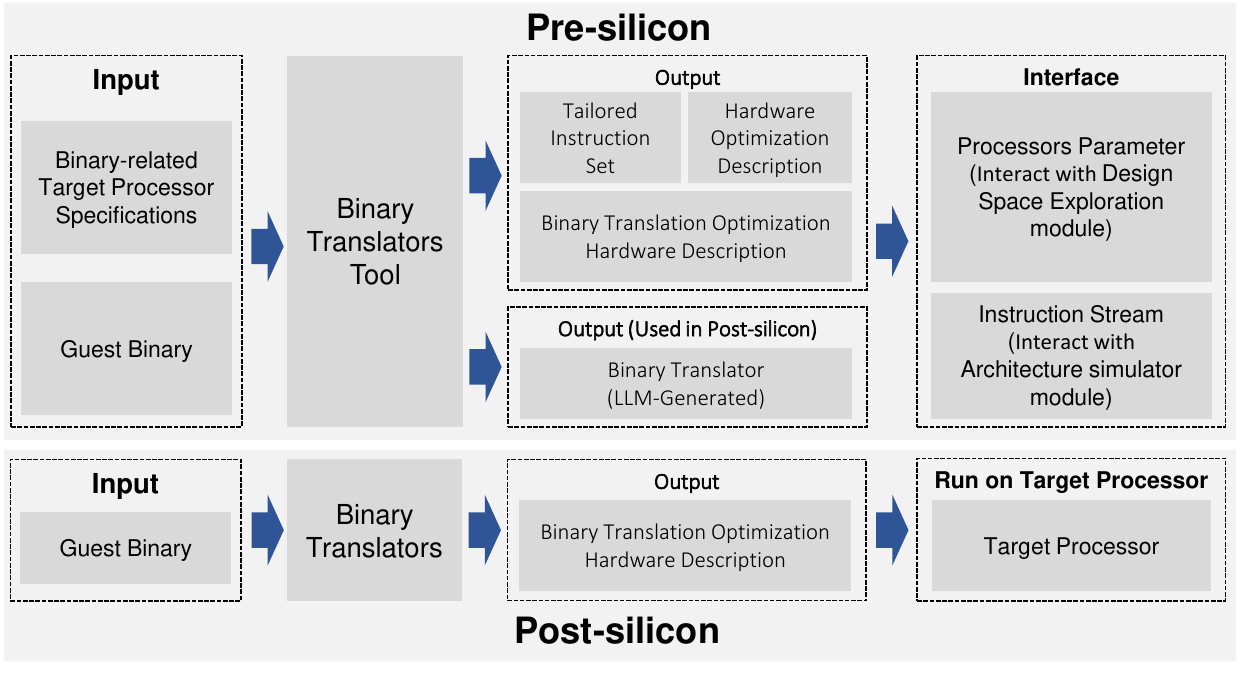}
    \caption{The hierarchical workflow of the LLM-driven binary translation module within LPCM. }
    \label{fig:4_2_flow_of_bin}
\end{figure}

Our work aims to build an advanced binary translation tool that intervenes as early as the pre-silicon design phase to introduce a paradigm shift, as illustrated in Figure~\ref{fig:4_2_flow_of_bin}. The primary objective is to automate the generation of binary translators, thereby significantly reducing or even eliminating the need for manual development. Tailored instruction sets, Hardware optimization descriptions and Binary translation optimization hardware description interact with other modules, while the tool continuously acquiring the latest processing data and performance evaluation results to support the iterative development of the advanced binary translation tool itself. In line with the three levels of LPCM development, the output process of the binary translation tool is also divided into three iterative phases within the workflow.

In the post-silicon verification phase, where the target processor design has been finalized, the binary translation tool shifts its focus to the construction of the runtime environment. It facilitates the stable and efficient execution of applications and dependent libraries originally developed for other ISAs or environments on the finalized hardware. Specifically, the binary translation tool supports rapid application porting and deployment, ensures backward compatibility, continuously optimizes hardware performance, and contributes to building a comprehensive software ecosystem for the new hardware. This comprehensive approach ensures that both hardware and software are fully integrated to maximize performance and usability across diverse computing environments.

In summary, the binary translation module serves as a crucial enabler for cross-ISA application migration, hardware/software co-optimization, and the intelligent evolution of system architectures within LPCM, ultimately bridging the gap between emerging hardware and rich software ecosystems.

%% file: submit_version/6_Simulator.tex
\section{Simulator meets LLM}
\subsection{Motivation}
Computer architecture simulators are software tools that model the structure and performance of computer systems or components such as CPUs and memory. As semiconductor technology advances, chip architectures become increasingly complex, making these simulators essential for research and development. They facilitate experimental analysis, rapid design iteration, and cost minimization before finalizing designs. However, as research in domain-specific acceleration evolves rapidly, the traditional approach to developing these simulators, which relies on human expertise, struggles to keep up with the pace of innovation. This development paradigm faces several significant challenges:

(1) \textbf{High Learning Curve}. Mature simulators like GEM5\cite{gem5,gem5v20}, booksim\cite{booksim}, and QEMU\cite{qemu} have developed extensive and complex codebases over the past decade, posing challenges for new developers. Understanding their architectural design and implementation can require 2\~4 weeks, hindering efficient modifications and research in hardware architecture. %With GEM5 alone comprising 30 million lines of code, there is a pressing need for innovative methods to simplify the configuration, usage, and development of these simulators, ultimately enhancing research and development efficiency in computer system architectures.

(2) \textbf{Simulator Composition Challenges}. Computer system architecture comprises several modules, including the CPU, GPU, NPU, DRAM, and SSD. However, existing simulators typically focus on individual components and employ distinct simulation methods and communication interfaces. This complexity can hinder development efficiency, underscoring the need for an automated integration scheme to streamline the process.

(3) \textbf{Prolonged Development Cycles}: The construction of traditional simulators typically involves multi-level abstract modeling, which spans from ISA functional validation to micro-architecture timing simulation, and is generally carried out manually. However, as the complexity of computer architecture system design continues to grow, the conventional manual implementation of simulators faces challenges such as tedious programming, intricate verification issues, and a time-consuming design space exploration process. These factors hinder the swift iteration of computer architecture system design.

Fortunately, the capabilities of large language models, particularly their advanced functionalities in code generation, code analysis, and debugging, offer an opportunity to address the aforementioned challenges and reform the development paradigm of computer architecture simulators. Thereby, it is time to leverage LLMs to design an efficient framework for the automated generation of computer architecture simulators, enabling automated transformation of user requirements or research papers into computer system architecture simulator. The input and output of an automated design framework for computer system architecture simulator should have the following functions: 

%To establish an effective automated design framework for computer system architecture simulators, it should encompass the following essential characteristics: (1) comprehensive automation that facilitates the integration of simulator resources while optimizing parameters according to user requirements. This capability should also promote efficient interconnections through standardized protocols, thereby enhancing scalability for large-scale simulations. (2) a dynamic equilibrium between simulation accuracy and speed, allowing for flexible adjustments based on scenario requirements. This can be achieved through selecting varying simulation techniques, ultimately improving both efficiency and accuracy throughout the development cycle. Based on the above characteristics, the input and output of an automated design framework for computer system architecture simulator should have the following functions:

\textbf{Input}: The framework primarily accepts simulation configurations and simulated workloads as input. These configurations can be represented in various modalities, including but not limited to natural language, documents, tables, and structured formats such as JSON and XML. The simulated workloads are generated by the upstream compiler module and typically appear in binary form or as user-defined intermediate representations.

\textbf{Output}: The output of the framework mainly includes two parts: complete simulator project codebase and user-required performance metrics \& analysis results. The former is the engineering code corresponding to the generated architecture simulator, including source code, documentation, etc., provided to users for code verification or secondary modification, while the latter is the simulation results required by users to verify the effectiveness of the design and achieve design optimization iteration.

%The framework's output must not only maintain functional equivalence with traditional simulators, but also leverage the rapid iterative modification capabilities of LLM. Accordingly, the output should include: simulator source code, simulation PPA, and interactive interfaces. (1) Simulator Source Code: The generated codes should include comprehensive code files, README documentation, and other necessary components to ensure the code executability and performance accuracy. Output simulation results and key metrics such as PPA (Power, Performance, Area) are also necessary. (2) PPA Reports: The framework can provide  (3) Interactive Interfaces: An interactive framework allows users to make rapid adjustments and optimizations based on preliminary simulation results, just like LLM web UI, thereby further advancing the efficiency of the development process. 

%Additionally, when incorporated into the overall architecture workflow, the simulator module receives simulatable code from the compiler module as its simulation workloads. Upon successful completion of the simulation, the simulator generates critical outputs, including configuration parameters, dataflow, specifications, and other essential data for the RTL module. Ideally, during normal operation, the output of the simulator should align with that of the RTL module.

%Through the above design, this framework would not only significantly shorten the development cycle and reduce costs but also markedly improve iteration efficiency, bringing transformative advancements to the field of chip development and design.

\subsection{LLM-driven computer architecture simulator automation design framework}

\begin{figure}
    \centering
    \includegraphics[width=1\linewidth]{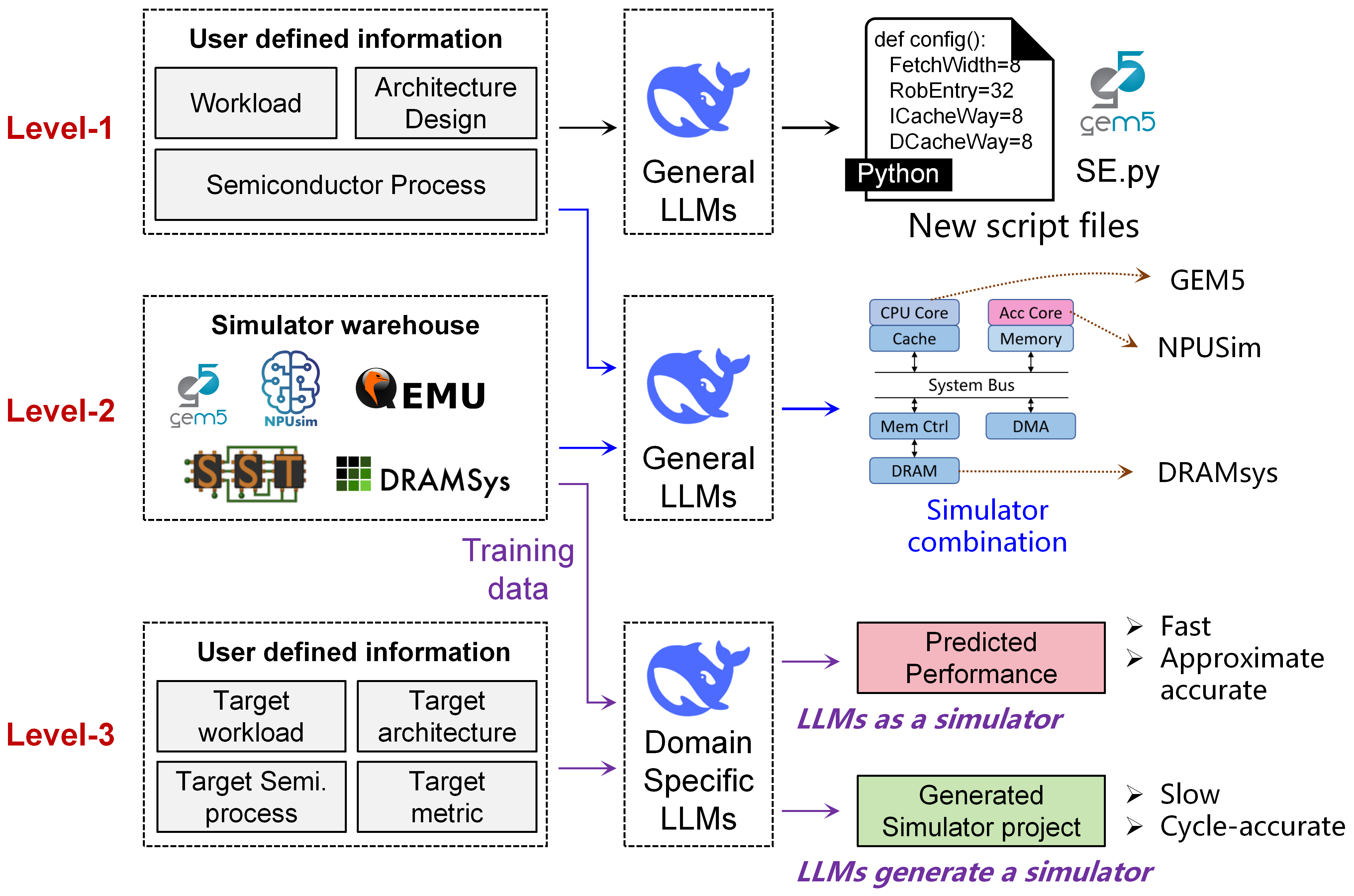}
    \caption{The three levels of design automation for LLM-driven computer system architecture simulators.}
    \label{fig:5_2_simulator_level}
\end{figure}

Despite the remarkable progress in LLM technology in recent years, developing an LLM-driven automated design platform for computer system architecture simulators with comprehensive end-to-end capabilities remains a significant challenge that cannot be overcome in a short time. In this context, this paper presents an automation grading standard specifically for the design of computer system architecture simulators powered by LLMs. As shown in Figure~\ref{fig:5_2_simulator_level}, the automated design process is organized into three levels, categorized according to the dimensions of functional realization and scenario requirements.
    
\begin{itemize}
    \item \textbf{Level 1: Intelligent Configuration of Simulators Based on General-Purpose LLMs}. The objective of Level 1 is to simplify and alleviate the cumbersome process of manually configuring simulators by harnessing the capabilities of advanced LLMs. By developing an efficient and intelligent mechanism that automates the conversion of natural language commands into precise simulator configurations, researchers can shift their focus away from the tedious aspects of simulator setup, allowing them to dedicate more energy and creativity to the computer system architecture design.
    
    %At the L1 stage, for a given architecture simulator development task with specified dataflow and architecture, there are typically two approaches. The first approach involves direct development of simulator code, requiring human programmers to complete tasks such as code module partitioning, unit testing, and fault debugging. The second approach considers utilizing existing configurable domain-general simulators, such as Gem5 or DRAMSim. For developers, writing code from scratch is not only time-consuming but also demands significant effort in debugging when the task requirements can be met otherwise. Therefore, most developers prioritize using existing simulators by configuring their parameters to fulfill simulation needs. Similar to developers' practices, the objective at the L1 stage is to leverage general-purpose large language models (LLMs) for intelligent parameter configuration of simulators rather than direct code generation. Specifically, by employing general-purpose LLMs, configuration requirements from users are formulated as prompts to achieve automated and intelligent parameter configuration for domain-general simulators. Subsequently, through automated workflows or agent mechanisms, the automated operation of simulators can be realized, thereby enabling further tasks such as automated design space exploration.
    
    \item \textbf{Level 2: Agent-Driven Simulator Composition}. The L1 level is dedicated to the automatic configuration of parameters for a single simulator, whereas the L2 level builds on this foundation by focusing on the automated interconnection of multiple simulators. The primary goal of Level 2 is to utilize large language models (LLMs) to bridge gaps among existing simulators in terms of simulation mechanisms, communication protocols, and interface standards. Attaining Level 2 has the potential to significantly reduce the time and effort researchers invest in simulator integration, while also improving the overall efficiency and flexibility of simulator design.
    
    %For the L2 stage, considering the current limitations of LLMs in modifying code at the codebase level, we propose leveraging them to compose and integrate simulators with different functionalities. This approach enables the development of simulators with novel architectures, and dataflow patterns. Taking an NPU(Neural Processing Unit) simulator as an example, its core modules typically consist of a computing array and on-chip memory. Different computing arrays exhibit distinct dataflow patterns and architectures, while variations in on-chip memory reflect differences in memory techniques. At the L2 stage, we focus on combining these components to develop NPU simulators with diverse architectures. In this phase, the LLM, assisted by AI agents, a simulator composition framework, and RAG techniques, performs modular encapsulation and integration. The final output is a simulator tailored to meet user requirements.
    
    \item \textbf{Level 3: Autonomous Simulator Design, Generation, and Optimization}. By utilizing the vast and varied data resources collected during Level 2, such as simulator source codes, workloads, and dynamic performance benchmark datasets, we can create domain-specific LLMs tailored for simulators. This progress has the potential to transform the design approach of computer system simulators, moving from a human-centered methodology to complete process automation.
    
    %At the L3 stage, with advancements in LLM technology, domain finetuned LLMs are expected to achieve codebase-level code generation and modification. In this phase, domain-specific LLMs can generate simulator module code based on user-defined requirements such as dataflow patterns, architecture specifications, and input/output interfaces. Building upon the generated simulator modules and existing simulators, the LLM and its associated agents can still perform L2-stage functionalities, composing and integrating these modules, to ultimately produce a complete architectural simulator. Concurrently, the L3 stage retains the capabilities of the L1 stage. For simpler architectural simulation requirements, the LLM will first leverage agent-based retrieval mechanisms to determine whether configurable general-purpose simulators can meet the user's needs. Only if this approach proves insufficient will the system proceed with the end-to-end simulator code generation pipeline described above.
\end{itemize}

    %Currently, while most simulators rely on manually written code, certain simulators are well-suited to serve as domain-general simulators in the L1 stage. Specifically, in the CPU domain, Gem5 enables simulations with different configurations through Python script customization. In the memory domain, DRAMSim and Ramulator provide various memory architectures and configurations, allowing users to specify different configuration files for simulations. In the NPU domain, frameworks like ONNXim and Stonne also support parameter configuration to some extent. These simulators can serve as domain-general solutions and be automatically configured during the L1 stage. At the L2 and even L3 stages, while there is currently no work focused on LLM-based automated simulator design, some research has been conducted on developing frameworks for simulator composition. In this regard, SST provides a unified API interface that enables interconnection between simulators after they are integrated into the framework. Another effort, LegoSim, developed a network for interconnecting different chips, where communication occurs through multiple threads and file interactions to achieve co-simulation of different simulators. Although these frameworks cannot be directly applied at the L2 stage, we can build upon them and leverage large language models to accomplish the task of simulator composition.

We have developed and implemented an intelligent computer system architecture simulator configuration solution leveraging the capabilities of a general-purpose LLMs, successfully accomplishing the core objectives of the L1 stage. In our experiments within the CPU domain, we selected the DeepSeek-V3\cite{simulator-dsv3} as the foundational model to configure GEM5\cite{gem5} simulator. By processing user-defined parameter configuration requirements, we effectively generated customized $SE.py$ simulation script files that meets user requirements, utilizing the unmodified $SE.py$ file as the starting template. Meanwhile, the generated scripts successfully passed the workload tests for 3D Gaussian Splatting\cite{simulator-3dgs} (3DGS), confirming their validity and accuracy. For the NPU domain experiments, we utilized the DeepSeek-V3 model, without any fine-tuning, and employed partial systolic array modules from the Gem5-Aladdin\cite{gem5-aladdin} simulator as configuration templates. With minimal human feedback for adjustments, we successfully created an NPU simulator specifically optimized for the 3DGS workload, based on the configuration parameters provided by the user. After integrating the generated code segments into the Gem5-Aladdin project, the simulator successfully passed the GEMV and GEMM tests within the workload, further establishing its performance and reliability. In our end-to-end 3DGS workload simulation experiments, we conducted comprehensive simulations of the 3DGS algorithm using the generated CPU and NPU modules. For the CPU module, we automated the complete parameter search process by integrating it with a design space exploration module, evaluating eight different parameter configurations before identifying the optimal set. Similarly, for the NPU module, we conducted an in-depth exploration of six key parameters, including systolic array dimensions and SRAM size, through the design space exploration module, ultimately determining the best configuration. Subsequently, the generated modules are integrated into the Gem5-Aladdin framework, facilitating inter-module communication through shared memory to support the accelerated execution of the 3DGS algorithm. The experiments contrasted CPU-only mode with CPU-NPU mode for simulating 3DGS. The results revealed that the CPU-NPU simulator achieved over a 20\% improvement in end-to-end performance compared to CPU-only execution for the 3DGS algorithm. This outcome not only confirms the effectiveness of our solution in automating simulator configuration but also highlights its considerable potential for enhancing algorithm execution performance.

In future work, we will further explore using LLMs to achieve Level 2 and Level 3 automated computer architecture simulator generation. Specifically, to achieve efficient integration of L2 level simulators, we will first develop a standardized interface protocol for simulator interconnection. This will enable plug-and-play connectivity among simulators using different simulation mechanisms by defining a unified data exchange format and communication specifications, thereby avoiding the technical bottleneck associated with extensive code refactoring in traditional integration approaches. Secondly, we will establish a comprehensive knowledge base of computer system architecture. This knowledge base will include two main types of data: the first encompasses simulator engineering data, systematically featuring the code bases of prominent architectures such as GEM5\cite{gem5} and Booksim\cite{booksim}; the second comprises knowledge on simulator interconnection and design methodologies. By leveraging the capabilities of LLMs alongside this knowledge base, the system will be equipped to semantically parse user requirements, identify the optimal simulators from the knowledge base, and seamlessly facilitate efficient interconnection among different simulators through the use of standardized interface protocols.

To achieve the goal of automating the design and generation of architecture simulators without human intervention at the level 3, we leverage the impressive capabilities of large models in code generation and performance prediction. Our approach consists of two distinct technical routes, focusing on the core dimensions of simulator accuracy and speed. 

(1) \textbf{LLM as an simulator}. By gathering workload code, hardware architecture details, and performance data from code executed on various architectures, we engage in domain-specific LLMs for computer architecture system simulator. This model is capable of accurately analyzing different workloads and hardware design schemes, simulating the execution process of workloads on various hardware architectures, and generating precise user required performance evaluation results. Essentially, the LLMs evolve into a simulator that incorporates performance prediction capabilities, making it suitable for architecture simulation and performance evaluation.

(2) \textbf{LLM generates an simulator}. The distinctive strengths of LLMs in code generation offer us a remarkable opportunity to translate researchers' design requirements into computer architecture system engineering code. In this way, researchers need only to articulate their design specifications, targeted performance metrics, and other critical information in natural language. The LLMs can then automatically interpret these requirements, transforming them into precise and standardized computer architecture system engineering code through their advanced semantic understanding and coding capabilities.

%% file: submit_version/7_hw_sw_partition.tex
\section{HW/SW partition meets LLM}

\subsection{Motivation}
% 软硬件划分的重要性、面临的挑战：优化周期长、对经验和专业知识依赖程度高、手动优化可能难以充分优化
As heterogeneous computing systems grow increasingly complex, hardware-software partitioning has become an indispensable step in the development process. Hardware-software partitioning is essential for rapid validation and iteration. At the early stages of system development, partitioning helps designers quickly establish system models, estimate performance, and analyze bottlenecks, providing clear guidance for subsequent hardware design and software development. By effectively distributing tasks between software and hardware modules, it not only optimizes performance and power consumption but also maximizes resource utilization. However, traditional partitioning methods face the following major challenges:
\begin{enumerate}
    \item \textbf{Low development efficiency and lengthy optimization cycles}: Hardware-software partitioning involves complex task decomposition, hardware resource mapping, and interaction logic analysis. Manual tuning is time-consuming and struggles to meet the demands of rapid iteration.
    \item \textbf{High dependency on expertise}: Partitioning requires developers to possess deep expertise in both hardware architecture and software design, increasing human resource costs and potentially leading to suboptimal results when expertise is lacking.
    \item \textbf{Underutilized optimization opportunities}: Without systematic learning from historical design patterns and large-scale data, manual partitioning often fails to uncover hidden performance optimization potential.
\end{enumerate}

% 引出LLM for HW/SW partition，说明理想框架的输入输出
To address these challenges, a highly efficient framework leveraging LLMs for hardware-software partitioning automation can be designed to support end-to-end optimization, from requirement specification to partitioning scheme generation. The ideal framework’s input and output paradigms are as follows:

\textbf{Input}: The framework should support diverse input formats, including natural language descriptions of requirements and original task C/C++ code. To enable flexible customization, it should also allow users to specify hardware resource parameters and partitioning constraints via structured formats such as JSON.

\textbf{Output}: (1) Partitioning scheme: The framework generates detailed partitioning schemes tailored to specific hardware architectures, specifying the allocation of each task to software or hardware.
(2) Performance prediction report: Outputs key performance metrics such as power consumption, performance, and latency to help users quickly evaluate the scheme’s quality.

With this design, the framework can significantly reduce development cycles and optimization time while uncovering hidden performance improvement opportunities through large-scale data learning. This dual enhancement of development efficiency and design quality represents a transformative advancement in the field of hardware-software co-design.

\subsection{Proposed framework design}
Based on the performance and involvement of LLM we proposed above, there are three progress challenging design approaches:

% 介绍L1-L3的软硬件划分模式
\begin{itemize}
    \item \textbf{Level 1: Human-guided HW/SW Partitioning}. At this level, human engineers remain the primary decision-makers, while LLMs serve as intelligent assistants. Leveraging their capabilities in natural language understanding and code analysis, LLMs can provide concise summaries, interpret complex modules, and analyze memory access patterns. These insights help engineers make informed design decisions more efficiently, although the final partitioning choices are still made by humans. %LLM acts as an assistant tool, analyzing task characteristics such as compute intensity and data dependency to recommend coarse-grained hardware/software partitioning strategies and generates interface code templates. However, the main tasks, such as task decomposition, task graph construction and performance evaluation, still require human efforts based on some traditional methods. Traditional methods typically rely on functional decomposition and task modularization, combined with analysis of real-time constraints, parallelism, computational load, and power consumption, to decide task placement.
    \item \textbf{Level 2: Agent-Aissisted HW/SW Partitioning}. At this level, LLMs take on a more proactive role. Acting as intelligent agents, they can parse source code, generate task graphs, and suggest preliminary partitioning strategies based on computational characteristics and platform constraints. Through iterative feedback from users, these agents can refine their recommendations, enabling a co-creative workflow between humans and machines. % LLM constructs fine-grained task graphs and dynamically optimizes subtask mapping (e.g., assigning convolutional layers to NPU while offloading loops to CPU) based on real-time hardware states (e.g., FPGA resource availability, memory bandwidth). It employs reinforcement learning to balance objectives like latency and power, enabling cross-layer adaptive partitioning.
    \item \textbf{Level 3: LLM-Driven Autonomous HW/SW Partitioning \& Optimization}. At this level, LLMs are deeply integrated with compilers, performance models, and learning frameworks such as graph neural networks. They are capable of executing the entire pipeline: analyzing code, constructing task graphs, estimating computational and memory profiles, and conducting multi-objective design space exploration to identify Pareto-optimal partitioning schemes. Moreover, LLMs can automatically generate the necessary software-hardware interface code, facilitating seamless deployment. %LLM autonomously decomposes tasks and designs architectures, generating custom hardware accelerators (e.g., specialized pipelines for sparse computations) and restructuring software stacks. It verifies partition consistency via formal methods but faces trust challenges due to opaque decision-making.
\end{itemize}

Most current approaches to HW/SW partitioning still rely on traditional methods, with limited involvement of LLMs. These conventional techniques typically depend on combinatorial optimization, heuristic algorithms \cite{eles1997system} \cite{lo1988heuristic}, and performance estimation models to determine task allocation. Heuristic-based methods—such as genetic algorithms\cite{holland1992genetic} \cite{zou2004hw}, simulated annealing\cite{van1987simulated}, and hill climbing\cite{selman2006hill}—aim to iteratively explore the solution space in search of optimal or near-optimal partitioning strategies under constraints like performance, power consumption, and hardware resource utilization.

According to the input and output requirements outlined in Section 1.1, the LLM-based automated hardware-software partitioning framework must fulfill two core functions: (1) Decomposition of Complex Tasks: Automatically analyze and decompose the input program into multiple simpler subtasks and generate the corresponding task dependency graph. (2) Rapid Task Performance Estimation and Architecture Mapping: Quickly estimate the performance and resource consumption of individual nodes and the entire task graph, enabling predictions of overall PPA (Power, Performance, and Area) under different hardware-software partitioning schemes. 

\begin{figure}
\centering
\includegraphics[width=1\linewidth]{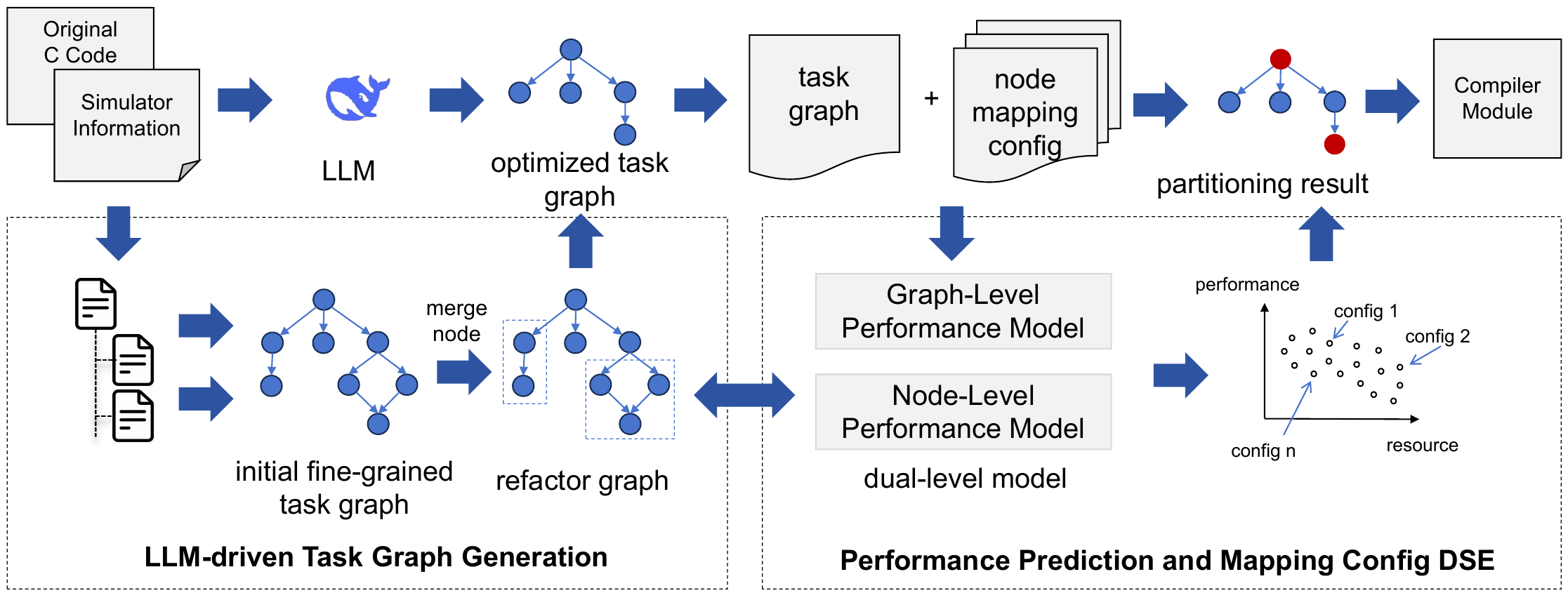}
\caption{Automated hardware-software partitioning framework}
\label{fig:6_overall_workflow}
\end{figure}

Based on the considerations above, we propose a method combining LLMs and graph learning models to achieve these key functionalities. The overall workflow of the proposed automated hardware-software partitioning framework is shown in Figure~\ref{fig:6_overall_workflow}. The framework primarily consists of two components:
(1) \textbf{LLM-Driven Task Graph Generation}: This component automates the decomposition of input programs and generates the corresponding task dependency graph. The goal of decomposition is to transform the input program from a complex and tightly coupled structure into a more transparent and interpretable task graph data structure. In the task dependency graph, nodes represent different architectural tasks, while the dependencies between these tasks are captured by the edges of the graph. This approach reformulates the hardware-software partitioning problem into a problem of managing dependencies between nodes in the task dependency graph, enabling more efficient and actionable task scheduling and partitioning in subsequent steps. (2) \textbf{Graph Learning Model-Based Task Graph Performance Prediction and Partitioning}: This component employs a dual-layer performance prediction framework powered by graph learning models to rapidly assess the performance and resource consumption of individual nodes in the task dependency graph, as well as the overall performance of the task graph under different mapping configurations. This framework facilitates efficient exploration of the design space, enabling the identification of the optimal hardware-software partitioning scheme under given architectural and resource constraints, thereby optimizing system performance and resource utilization. 

\subsubsection{LLM-Driven Task Graph Generation}
    % 作为软硬件划分的第一步，首先通过大模型agent来对输入的C/C++代码进行语义分析。大模型agent具备深入理解代码结构的能力，并将程序中的多个功能模块划分为更细粒度的子任务。通过分析程序的语义和整体结构，agent能够识别出子任务之间的依赖关系，并基于这些信息生成一个包含任务描述、依赖节点以及原始代码的json格式任务依赖图。这种图结构不仅能够清晰地展示各个子任务之间的执行顺序和依赖关系，还能为后续的优化和任务节点架构分配提供必要的信息。
    As the first process of HW/SW partitioning, this step leverages LLM to generate an appropriate task dependency graph from the original C/C++ source code provided by the user. It begins by using an LLM to produce an initial task graph with the finest granularity. In this initial graph, each function in the input C/C++ code is treated as a separate task node, and edges are constructed based on function calls and data dependencies. During this phase, the LLM analyzes the semantics and structural relationships of the source code to generate nodes and infer the initial dependency edges.
    Next, LLM performs graph optimization through node merging. It iteratively examines all directly connected node pairs and evaluates the benefit of merging each pair, aiming to balance execution time and communication overhead. The merging benefit is estimated based on the predicted performance characteristics of the nodes and the volume of data exchanged between them. After evaluating all candidate pairs, LLM selects the one with the highest gain for merging, updates the task graph accordingly. This process repeats iteratively until no further beneficial merges can be performed. Through the above methodology, LLM generates a task dependency graph that balances both execution time and communication overhead. The final output is a structured JSON representation of the task graph, produced by the LLM. %this step begins with an LLM agent performing semantic analysis of the input C/C++ code. The agent, equipped with an in-depth understanding of code structures, divides the program's functional modules into finer-grained subtasks. By analyzing the program's semantics and overall structure, the agent identifies dependencies between subtasks and generates a task dependency graph in JSON format. This graph includes task descriptions, dependency nodes, and the original code. The resulting graph structure clearly illustrates the execution order and dependencies among subtasks, providing essential information for subsequent optimization and architectural allocation of task nodes. The workflow of task decomposition is shown in Figure~\ref{fig:6_task_decomposition}.

    \begin{figure}
    \centering
    \includegraphics[width=0.7\linewidth]{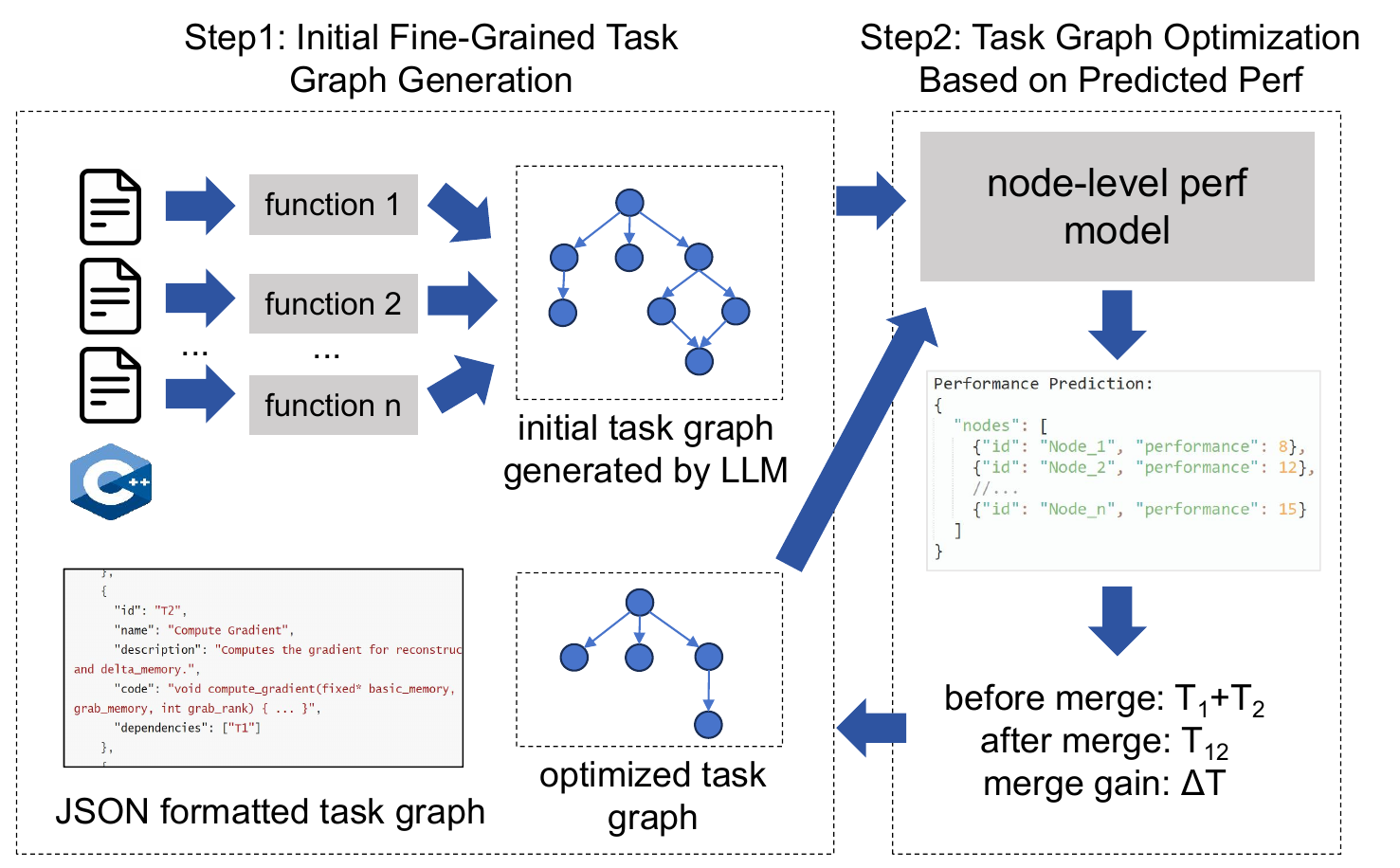}
    \caption{Workflow of task graph generation}
    \label{fig:6_task_decomposition}
    \end{figure}

    % 为了更好地对任务依赖图进行性能建模，随后对各个节点进行初步的性能分析。该步骤的核心在于评估那些难以通过传统性能分析工具量化的指标，比如代码的并行化潜力、逻辑复杂性、内存访问密集度等。为了全面捕捉任务的性能特征，这些特征被划分为多个类别，包括计算特征、访存特征、代码结构特征以及I/O特征等。通过大模型agent，结合代码本身的内容和编译器生成的数据流图（DFG），对各个子任务的这些特征进行标注，并将它们作为任务图节点的属性。这一过程使得任务依赖图不仅仅是一个静态的结构，而是包含了丰富的性能信息，为后续进一步的性能建模提供精准的依据。
    % To model the performance of the task dependency graph, an initial performance analysis is conducted for each node. The core of this step lies in evaluating metrics that are challenging to quantify using traditional performance analysis tools, such as the parallelization potential, logical complexity, and memory access intensity of the code. To comprehensively capture the performance characteristics of tasks, these features are categorized into multiple dimensions, including computational features, memory access features, code structure features, and I/O features.
    % Using the LLM agent, in combination with the program's content and compiler-generated data flow graph (DFG), these features are annotated for each subtask and incorporated as attributes of the nodes in the task graph. This process transforms the task dependency graph from a static structure into one enriched with detailed performance information, laying a solid foundation for further performance modeling and optimization.

\subsubsection{Graph Learning Model-Based Task Graph Performance Prediction and Partitioning}
\begin{figure}
    \centering
    \includegraphics[width=1\linewidth]{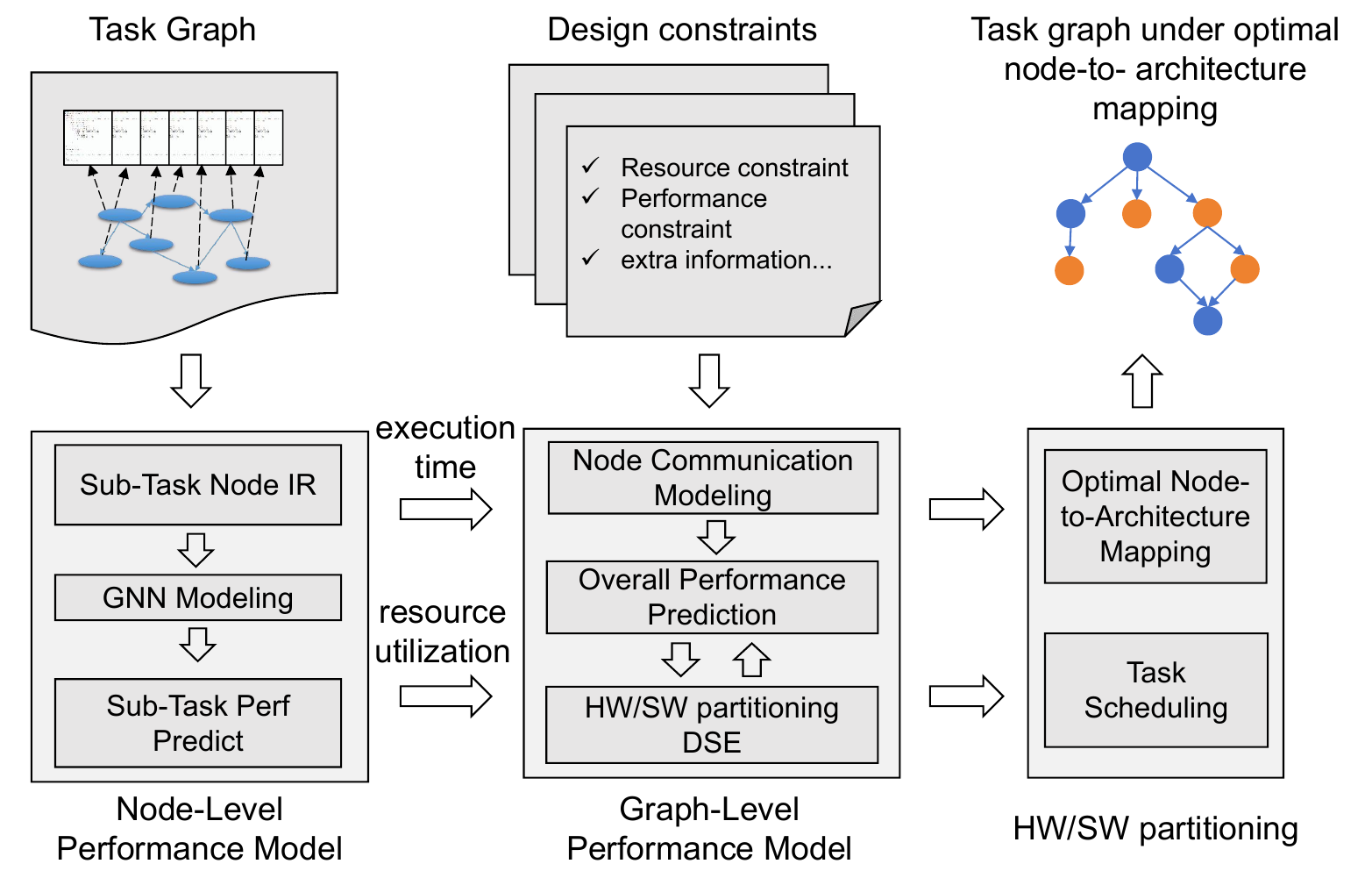}
    \caption{Dual-layer performance prediction.}
    \label{fig:6_performance_model}
    \end{figure}

    % 该部分工作基于一种利用图学习模型的双层性能预测模型，旨在快速预估任务依赖图中各个节点的性能与资源消耗，同时预测整个任务图在不同节点-架构映射配置下的整体性能。通过这一框架，可以实现对设计空间的高效探索，从而在给定架构和资源约束下找到最优的软硬件划分方案，优化系统性能和资源利用率。
    This work is based on a dual-layer performance prediction model utilizing graph learning techniques. The goal is to efficiently estimate the performance and resource consumption of each node in the task dependency graph while predicting the overall performance of the entire task graph under different node-to-architecture mapping configurations rapidly. This framework enables effective design space exploration, allowing the identification of optimal software-hardware partitioning solutions within given architectural and resource constraints, ultimately optimizing system performance and resource utilization. The workflow of performance prediction is shown in Figure~\ref{fig:6_performance_model}.

    % 性能预测模型采用双层设计，分别为节点级性能预测模型（node-level）和图级性能预测模型（graph-level）。在节点级模型中，每个任务依赖图中的节点被视为一个独立的子任务。模型的输入为任务图中每个子任务对应的C/C++代码，输出该子任务在不同平台（CPU和FPGA）上的性能和资源消耗预测值。具体而言，首先通过高层次综合工具将子任务代码转化为中间表示（IR），这种表示形式可以有效抽象出代码的底层计算和访存特征，便于后续建模。随后，利用图神经网络对每个节点的性能和资源消耗进行建模，GNN的特性使得其能够有效捕捉任务节点的特征以及IR结构中的局部依赖关系，从而提高预测精度。
    
    The performance prediction model consists of two layers: a node-level performance prediction model and a graph-level performance prediction model. 
    In the node-level model, each node in the task dependency graph is treated as an independent subtask. The model takes as input the C/C++ code corresponding to each subtask and outputs predictions for the subtask's performance and resource consumption on different platforms (CPU and NPU). Specifically, the subtask code is first transformed into an intermediate representation (IR) using high-level synthesis tools. This IR effectively abstracts the underlying computational and memory access characteristics of the code, facilitating subsequent modeling. Next, a graph neural network (GNN) is employed to model the performance and resource consumption of each node. The GNN’s ability to capture local dependencies within the IR structure enhances the accuracy of the predictions by leveraging both the features of the task nodes and the structural relationships in the IR.
    
    % 图级性能预测模型则用于快速评估在特定节点-架构映射配置下的任务图整体性能。模型的输入包括整个任务依赖图、节点的架构映射配置，以及由节点级模型预测的每个子任务在不同平台上的性能数据，包括执行时间、资源消耗等。该部分模型综合考虑任务图中节点间通信数据、任务调度和资源竞争等关键因素，对任务图整体性能建模。
    The graph-level performance prediction model is designed to rapidly evaluate the overall performance of the task graph under specific node-to-architecture mapping configurations. The model’s input includes the entire task dependency graph, the architecture mapping configuration of the nodes, and performance data for each subtask as predicted by the node-level model, such as execution time and resource consumption. This model accounts for critical factors such as inter-node communication data, task scheduling, and resource contention within the task graph to construct a comprehensive performance model.
    
    % 基于图级模型的输出，可以快速评估任务依赖图在所有可行的节点-架构映射配置下的性能表现，结合提前给定的资源约束等信息，可实现设计空间探索，最终得出合理的软硬件划分方案和任务调度策略，确保系统整体设计最优。
    Based on the output of the graph-level model, we can quickly evaluate the performance of the task dependency graph across all feasible node-to-architecture mapping configurations. By combining this with predefined resource constraints, the framework enables efficient design space exploration. This process ultimately identifies a reasonable software-hardware partitioning scheme and task scheduling strategy that ensures optimal overall system design.

%% file: submit_version/8_dse.tex
\section{Design Space Exploration meets LLM}
\subsection{Motivation}
%\subsection{The Role and Workflow of DSE Module in The SAM}

Design Space Exploration (DSE) plays a vital role as a strategic intermediary between algorithmic requirements and hardware architecture, enabling the co-design of CPUs and co-processors under stringent performance, power, and area (PPA) constraints. As computational demands continue to diversify, CPUs must adapt through architectural refinement to maintain versatility, while co-processors leverage specialization to accelerate targeted workloads. Together, they push the boundaries of system-level performance. However, this architectural flexibility results in exponentially growing design spaces and increasingly complex parameter interactions. 

Traditional manual DSE processes are typically slow, requiring months or even years of iterative optimization, and have become a significant bottleneck in the hardware development lifecycle. This challenge has led researchers to pursue intelligent DSE techniques tailored for both CPUs and co-processors, aiming to rapidly identify optimal design points across vast configuration spaces. By automating this process, such methods promise to significantly reduce development time and foster faster hardware-software co-evolution.

%DSE must be automated to keep pace with the increasing complexity of modern system architectures, which involve vast and high-dimensional design spaces that are impractical to explore manually. 
%Traditional manual methods are not only time-consuming but also struggle to efficiently balance multiple, often conflicting objectives such as performance, power, and area. In contrast, automated DSE enables intelligent, data-driven optimization, rapidly adapts to diverse and evolving workloads, and significantly enhances both design efficiency and solution quality. As such, it serves as a key enabler for intelligent, scalable, and workload-aware hardware design.

However, achieving automated DSE comes with several key challenges. (1) Complex Integration within the System Architecture Flow: The DSE module must ensure consistency, compatibility, and accurate information exchange across various modules within the system architecture design workflow. Currently, significant manual intervention is still required to maintain coherence and correct functionality throughout the flow. (2) Offline Learning and Limited Adaptability:
As new architectures, technologies, and design methodologies continue to emerge, DSE modules must be frequently updated to remain effective. However, enabling robust online learning while avoiding catastrophic forgetting remains a significant challenge, limiting the system’s ability to adapt continuously and autonomously.

The integration of LLM into the DSE module can effectively addresses above challenges. LLM acts as a seamless coordinator across different modules, automatically understanding and transmitting design requirements and constraints, thereby reducing manual intervention and improving system consistency and efficiency. Additionally, LLM's robust knowledge generalization capability enables it to adapt to new architectures and technologies without the need for frequent retraining. By supporting incremental learning, LLM avoids catastrophic forgetting, ensuring continuous evolution. Through interactive optimization with designers, LLM fosters a dynamic, adaptive, and intelligent design process, significantly enhancing the automation and intelligence of the DSE module.

To address these challenges, we design a DSE module powered by LLMs. This module plays a foundational role in navigating the vast design space of modern processors. Within the LPCM framework, the primary objective is to autonomously generate and refine hardware designs that meet diverse functional, performance, and cost requirements—where both the CPU and co-processor are critical system components. The DSE module, in this context, refers to the structured and intelligent process of exploring and evaluating a wide range of CPU and co-processor configurations and architectures. Its goal is to identify the most optimal hardware design that satisfies the multi-dimensional constraints and objectives defined by LPCM.
The role of DSE module in the LPCM can be outlined as follows. 
\begin{itemize}
    \item \textbf{Automated Design Exploration.} DSE module automates the process of exploring numerous configurations of CPUs and co-processors, including cores, cache hierarchies, instruction sets, coprocessor architectures, extended instructions, cache coherence, and power/area/performance trade-offs. By systematically adjusting these parameters, DSE module identifies the configuration that best meets performance, power, area, and cost objectives.
    
    \item \textbf{Rapid Iteration and Optimization.} By leveraging performance predictions, different configurations are rapidly evaluated in an iterative manner, where feedback from each evaluation cycle informs subsequent design adjustments. This continuous refinement process significantly shortens development cycles and reduces reliance on costly and time-consuming trial-and-error methods.
        
    %\item \textbf{Self-refining Designs.} Beyond static exploration, CPU DSE within SAM is designed to adapt and refine CPU architectures continuously. The model’s intelligent agents leverage iterative feedback from simulations and real-world performance data to adjust design choices and optimize the architecture progressively, ensuring that each iteration moves closer to the ideal design.
     
    \item \textbf{Holistic System Optimization.} DSE module does not operate in isolation. Instead, it is part of a broader system-wide optimization process. The design of the CPU and co-processor is considered in conjunction with other system components such as compilers, software/hardware partitions, and memory subsystems. This interconnected approach ensures that the CPU and co-processor are optimized for efficient operation within the larger system, creating a fully integrated and optimized architecture.

\end{itemize}

\begin{figure}[!htbp]
    \centering
    \includegraphics[width=0.8\linewidth,page=1]{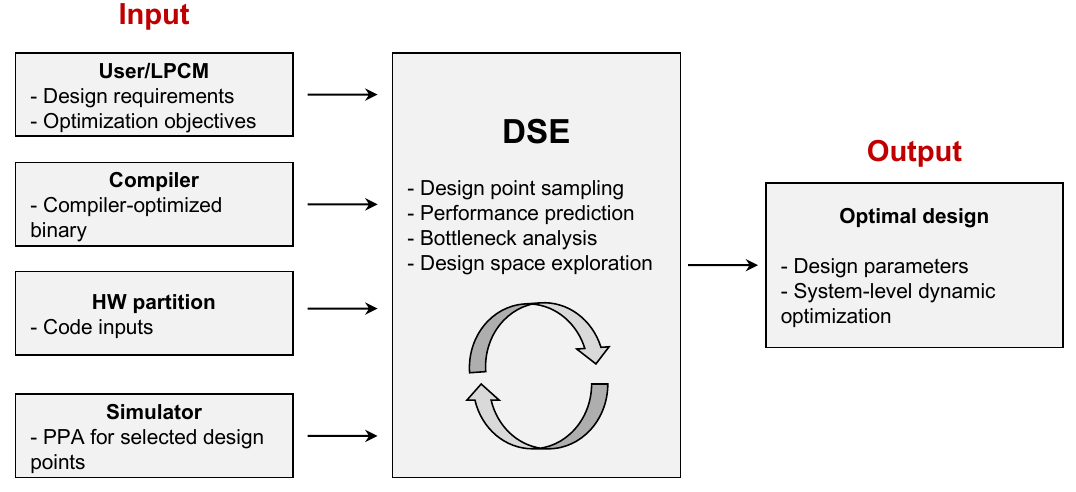}
    \caption{Workflow of CPU DSE module on LPCM.}
    \label{fig:CPU_DSE_workflow}
\end{figure}

The DSE module operates as illustrated in Figure~\ref{fig:CPU_DSE_workflow}. In the \textbf{Input}, it gathers essential inputs including design requirements and optimization objectives from the top-level LPCM, workload characterization code from the hardware-software partitioning module, compiler-optimized binaries for performance modeling, and PPA results of selected design points from the simulation module for training and validation purposes. During the exploration, the DSE module performs autonomous design point sampling with adaptive density control, predicts PPA metrics for unexplored configurations without requiring external simulation, and conducts bottleneck analysis to identify architectural inefficiencies—such as suboptimal cache hierarchy or pipeline depth—informing microarchitectural refinements. These capabilities are integrated to guide efficient and intelligent exploration of the design space toward optimal architecture configurations. Finally, in the \textbf{Output}, the DSE module delivers the optimized microarchitecture parameters—including core topology, execution units, and cache structures—to the RTL code generation backend. Simultaneously, LPCM leverages insights from the DSE process to perform system-level dynamic optimization, adjusting OS settings, compiler strategies, and hardware acceleration schemes, thus establishing a closed-loop co-optimization flow across the stack.

\subsection{The Overview of DSE Module}

%\subsubsection{Levels of Automation Definition for DSE Module}

%We propose the LLM DSE, an advanced design space exploration powered by LLMs, specifically designed for the prediction and optimization function within SAM. Based on the varying capabilities of LLMs and their level of involvement, we propose a tiered classification as follows, and illustrated in Figure~\ref{fig:CPU_DSE_workflow_various_level}:

\begin{figure}[!htbp]
    \centering
    \includegraphics[width=1\linewidth,page=2]{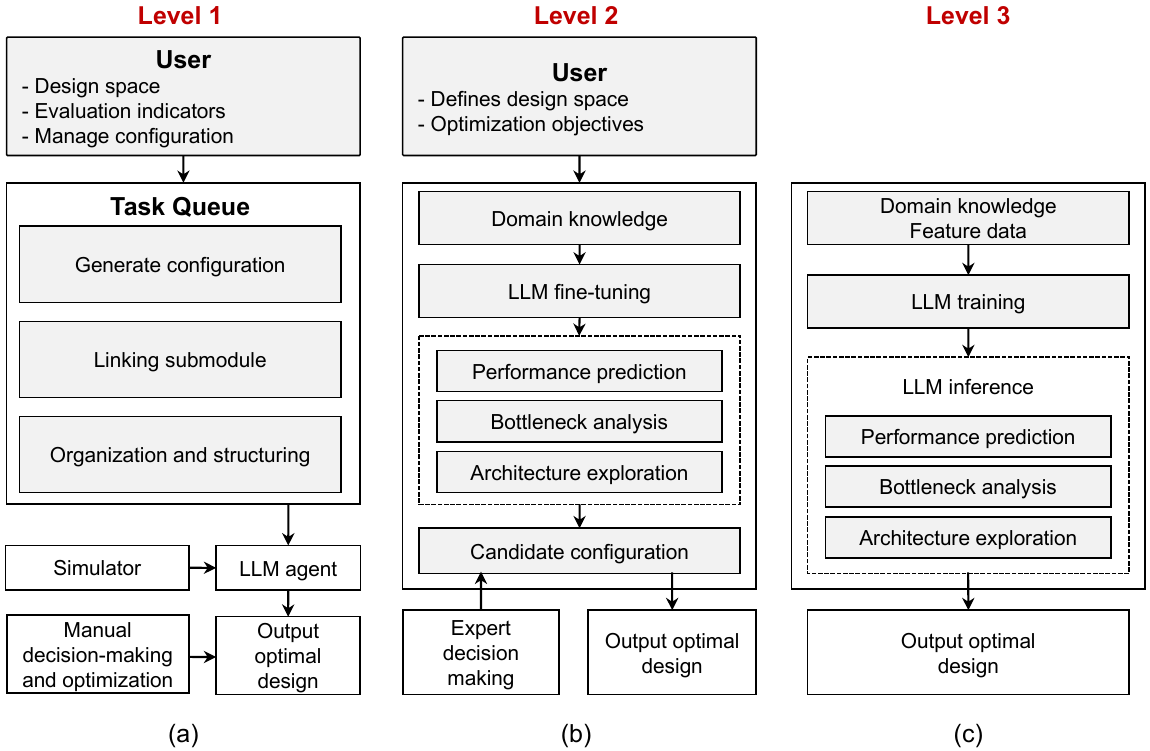}
    \caption{DSE module at (a) Level 1, (b) Level 2, and (c) Level 3.}
    \label{fig:CPU_DSE_workflow_various_level}
\end{figure}

To leverage the capabilities of LLMs and realize fully automated design space exploration without human intervention, we propose the LLM DSE, an advanced exploration framework powered by LLMs, specifically designed for prediction and optimization tasks within LPCM. However, achieving the goal of fully autonomous hardware design is a long-term endeavor that requires progressive technological advances. To support this objective, we define three levels based on the development trends and varying capabilities of LLMs, as illustrated in Figure~\ref{fig:CPU_DSE_workflow_various_level}.

\begin{itemize}
    \item \textbf{Level 1: Human-Guided and LLM-Assisted DSE}. At this level, LLMs function as intelligent assistants to streamline repetitive tasks such as organizing input data, linking evaluation tools, and generating basic configurations. While the core decision-making remains with human experts, LLMs contribute by chaining sub-modules and facilitating design flow integration. The system remains highly reliant on expert guidance for simulation execution, performance evaluation, and iterative refinement. LLMs support input preprocessing and workflow scripting, but do not participate in active exploration or optimization.
%Existing DSE frameworks~\cite{Boom_Explorer_21_ICCAD, MoDSE_TCAD_23, TrEnDSE_ICCAD, AttentionDSE_2024, MetaDSE_DAC_2025} focus on human-guided performance prediction models without LLM.

    \item \textbf{Level 2: LLM-Driven Semi-Automated DSE}. At this level, LLMs assume a more proactive role by autonomously generating candidate configurations, analyzing performance feedback, and iteratively refining design proposals under human supervision. Human experts define high-level objectives and constraints, while LLMs explore the design space by leveraging pre-trained knowledge and retrieval-augmented generation (RAG) techniques. The interaction becomes bidirectional—LLMs propose adjustments based on previous evaluation results, and experts respond by refining constraints or redefining targets. This level enables partial autonomy while retaining expert control over critical decision points within the design workflow.
    To effectively support diverse application scenarios, a general-purpose coprocessor DSE framework is required. This framework is typically built upon heuristic algorithms such as genetic algorithms, with LLMs embedded into key components to enhance adaptability and decision-making efficiency.
        
    \item \textbf{Level 3: LLM-Governed Fully Autonomous DSE}. At this level, LLMs govern the entire design space exploration process with minimal to no human intervention. The system autonomously synthesizes design inputs, generates and evaluates candidate architectures, and selects optimal configurations through internal reasoning and feedback-driven learning. It performs holistic, cross-layer exploration and supports multi-objective optimization across key metrics such as performance, power, and area. LLMs at this stage act as central decision-makers, orchestrating the full loop of configuration generation, simulation execution, result analysis, and iterative refinement, thereby achieving a high degree of autonomy in hardware design.  
\end{itemize}

Although recent efforts do not directly employ LLMs for DSE, they can serve as submodules within the DSE module. These submodules include design samplers, performance models, and exploration methods.
The design samplers encompass techniques such as random sampling~\cite{NN_06_ASPLOS}, orthogonal design~\cite{ActBoost_DAC_16, MoDSE_DAC_23}, and Pareto-aware sampling~\cite{MoDSE_TCAD_23}.
The performance models include linear regression~\cite{Linear_model_05_HPCA}, spline models~\cite{Spline_06_ASPLOS}, neural networks~\cite{NN_06_ASPLOS, ArchRanker_14_ISCA, ActBoost_DAC_16, AttentionDSE}, Gaussian processes~\cite{Boom_Explorer_21_ICCAD}, XGBoost regressors~\cite{Mcpat-calib_TCAD_23}, and GBRT regressors~\cite{MoEnDSE_GLSVLSI_23, MoDSE_DAC_23, MoDSE_TCAD_23}. Prediction accuracy can be further enhanced by leveraging transfer learning and meta learning techniques from known workloads~\cite{codes16_TrDSE, iccad17_TrEE, MPGM_ISAC_19, TrEnDSE_ICCAD_23, MetaDSE}. The exploration method includes heuristic search and acquisition functions. The heuristic search generates candidate design points using predefined rules, including random descent~\cite{eyerman_GA_2level_DATE_06}, genetic algorithm~(GA)~\cite{mariani_meta_model_2009, ML-NSGA-II_09, NGSA_ANN_ATECS_13, ACOSSO_ASPDAC_15, Multicube_ARCS_10}, and simulated annealing~\cite{DSE_critical_PACT_10, mariani_MOSA_08}.
%Panerati and Beltrame~\cite{Survey_DSE_cmp_ATDAES_14} provide a comprehensive comparison and classify these algorithms using various metrics.
The acquisition function relies on the statistical characteristics, including uncertainty~\cite{ActBoost_DAC_16}, 
expected improvement~\cite{mariani_correlation_based_DAC_10, mariani_OSCAR_TCAD_12}, and Pareto charateristics~\cite{Boom_Explorer_21_ICCAD, MoEnDSE_GLSVLSI_23, MoDSE_DAC_23, MoDSE_TCAD_23}.

Moreover, the design space of co-processor architectures is highly complex, and as task loads continue to diversify, it is crucial to thoroughly address the challenges of generality in co-processor design space exploration. The exploration of this design space can be classified into three primary levels based on the degree of generality in the methodologies employed. The first level focuses on optimization for single-task execution on specific co-processor architectures, with studies at this level typically involving the design of custom DNN accelerators~\cite{mei2021zigzag, hong2023dosa}. The second level deals with the optimization for general tasks on specific co-processor architectures, exemplified by automated design methods for systolic arrays~\cite{cong2018polysa, wang2021autosa}. The third and most advanced level concerns the automated selection and optimization of co-processor architectures for general tasks. Although research in this area is still limited, studies like~\cite{shao2014aladdin} have introduced methods for designing general-purpose co-processors, offering valuable insights for future co-processor design space exploration.

\subsubsection{Key Future Designs}

To achieve higher levels of automation, adaptability, and intelligence in processor design space exploration, future advancements must go beyond static decision-making pipelines. In this section, we present two key enablers that underpin the transition from human-guided design to fully autonomous systems: (1) a modularized reasoning and adaptive decision-making framework that empowers LLMs with structured, interpretable, and feedback-driven capabilities, and (2) a unified knowledge graph with online learning mechanisms to ensure continuous evolution and system-wide knowledge integration. Together, these foundational designs lay the groundwork for scalable, explainable, and continuously improving DSE processes across all levels of automation.

\textbf{Modularized Reasoning and Adaptive Decision-Making Framework.}
To support fully autonomous and reliable design space exploration at advanced automation levels, we propose a modular reasoning and adaptive decision-making framework that empowers LLMs with structured, verifiable, and context-aware inference capabilities. By decomposing the DSE process into modular sub-decisions—such as architectural parameter tuning, pipeline depth configuration, and power gating strategy selection—the LLM can apply localized heuristics and domain rules to each decision unit. This modular structure simplifies learning, supports hierarchical optimization, and improves reasoning interpretability. In parallel, the framework integrates adaptive reasoning mechanisms that dynamically adjust exploration strategies based on feedback signals such as constraint violations, PPA degradation, or convergence stagnation. Reasoning chains are established to trace the causal relationships between design choices and performance outcomes, enabling explainable inference and reduced decision uncertainty.
In the context of L1 to L3 levels of DSE module, we propose a modularized reasoning and adaptive decision-making framework to facilitate increasing autonomy and effectiveness at each level. 

 At L1, the modular reasoning framework plays a supporting role. LLMs assist human experts by organizing and structuring design parameters into modular units. For example, sub-decisions like architectural parameter tuning and pipeline configuration can be structured into modules, enabling the LLM to propose initial configurations. The modular approach simplifies the management of multiple design parameters, but human experts still provide final decisions. Adaptive decision-making here primarily supports experts by flagging potential issues (e.g., performance degradation or constraint violations) based on the modular structure, without taking full control.

At L2, the LLM can autonomously apply modular reasoning to evaluate and optimize design configurations. The framework will allow the LLM to perform more sophisticated analyses, such as dynamically adjusting strategies based on the feedback from previous iterations. For example, if a configuration violates constraints or results in PPA degradation, the LLM can adjust its exploration strategies by altering specific modules (e.g., adjusting power gating strategies or reconfiguring pipeline depth). The integration of adaptive reasoning ensures that the LLM can autonomously adjust the design space exploration process and make iterative improvements based on the feedback loop from simulation results.

At L3, the modular reasoning and adaptive decision-making framework allows the LLM to take complete control of the design space exploration process. The LLM independently generates and optimizes designs by making decisions across multiple modules and adapting its exploration strategy based on continuous feedback. Reasoning chains will enable the LLM to link design choices to performance outcomes, facilitating self-correction and improving design decisions autonomously. This results in fully optimized configurations and performance metrics without the need for human intervention.

\textbf{Online Learning with Unified Knowledge Graph Integration.}
To realize continuous adaptation and system-wide design intelligence, we introduce a unified knowledge graph framework coupled with online learning capabilities. The knowledge graph integrates heterogeneous data sources across abstraction layers, including microarchitectural configurations, RTL component libraries, EDA constraints, and empirical performance logs. Through a RAG-enabled interface, the LLM can retrieve relevant subgraphs in real time to inform design decisions, enabling contextual awareness and accurate cross-domain reasoning. Online learning pipelines are incorporated to incrementally refine the LLM's internal representation using newly acquired simulation results, verification outcomes, and emerging design patterns. This setup allows the system to autonomously evolve its design knowledge, adapt to novel hardware requirements, and remain aligned with the state of the art in CPU architecture and optimization techniques.
In line with the increasing autonomy at each level of DSE, we introduce an online learning framework that leverages a unified knowledge graph to enable continuous adaptation of the LLM. 

At L1, the knowledge graph serves as a central repository for organizing and referencing domain-specific information, such as component libraries and design constraints. The LLM can query the graph to suggest candidate solutions, but the expert remains in control of integrating new data. While online learning is limited at this stage, the system can assist in refining the decision-making process by providing relevant information and drawing from historical data for the experts.

At L2, the knowledge graph becomes more interactive, with the LLM retrieving real-time data from the graph to refine its design decisions. The integration of online learning allows the system to update its knowledge continuously based on simulation results and verification outcomes. As the LLM autonomously explores design configurations, it can incorporate new performance data, thereby improving its decision-making accuracy over time. The system becomes increasingly efficient in suggesting configurations that meet design goals based on updated knowledge and feedback.

At L3, the knowledge graph is fully integrated into the LLM’s decision-making process. The LLM can autonomously query the graph for relevant data and adjust its optimization strategy based on newly acquired insights. The online learning pipeline allows the system to evolve continuously, adapting to emerging design patterns and technological advancements. The LLM uses the updated knowledge base to drive its optimization process, making decisions based on the most current and relevant data available, ensuring that the design process stays aligned with cutting-edge developments.

%% file: submit_version/9_rtl.tex
\section{HDL generation module}

\subsection{Motivation}
The Hardware Description Language (HDL) generation module automatically creates executable HDL code based on the optimal design generated by the above Design Space Exploration Agent. It processes diverse inputs such as natural language specifications, C code, and control flow graphs to interpret design parameters and perform system-level dynamic optimization using a pre-trained model. The module includes an input interface, a multimodal HDL generator, and a PPA (Power, Performance, Area) prediction feedback mechanism to ensure functional correctness and performance optimization. 

As chip design complexity grows, integrating HDL generation into large processor chip models becomes essential—not only to enhance efficiency and reduce human errors, but also to achieve global optimization. This approach overcomes the limitations of traditional manual HDL writing while enabling cross-layer optimization, resulting in a more efficient and accurate design process. Below are key reasons why HDL generation should be embedded within such models.

\begin{itemize}
    \item Collaborative Optimization for Global Optimality

    Cross-layer collaborative optimization integrates HDL generation into system-level modeling frameworks, enabling seamless coordination between different design layers. By establishing closed-loop feedback among compilation optimizations, simulation tools, and design space exploration, this approach not only enhances design correctness and performance but also ensures component compatibility and consistency, driving toward a globally optimal solution.

    Consider the case of a complex microprocessor design. If HDL coding is done in isolation, it might result in under-use of available resources or suboptimal trade-offs between power consumption and performance due to ineffective communication with other design layers. However, by adopting a multimodal modeling approach at the system level, designers can make more informed decisions based on comprehensive data analysis, leading to an optimized overall design.

    \item Overcoming Limitations of Existing RTL Generation

    Traditional RTL code generation methods often suffer from incomplete information and insufficient validation feedback, making it difficult to guarantee the accuracy of the generated code. Without robust verification mechanisms, errors can propagate unchecked. In contrast, modern large-scale processor chip models incorporate end-to-end support, from compilation optimization to simulation and design space exploration, establishing a closed-loop feedback system that significantly improves design reliability and correctness.

    Handling highly complex hardware design requirements, such as architectural intricacies and resource constraints, poses significant challenges for traditional approaches, often leading to impractical designs or overly complex training and inference processes. Advanced large processor chip models offer a more effective solution by streamlining these complexities, enabling efficient automation of HDL generation while ensuring scalability and adaptability to diverse design constraints.
\end{itemize}

\subsection{Overview of HDL Generation Module}

\subsubsection{Development Stages}
We describe HDL generation across the following three levels. Currently, we are at level 3.

\begin{itemize}
    \item \textbf{Level 1: Prompt-centric Copilot}.
    In this initial stage, general LLMs are used to generate HDL code. Debugging prompts play a crucial role here. Previous works directly prompt LLMs to generate HDL code or explore prompt optimization using feedback from EDA tools.

    \item \textbf{Level 2: Fine-tuning Submodel}.
    In the fine-tuning stage, the HDL model is one of the submodels of LPCM. It has two key characteristics. 
    Large models fine-tuned on domain-specific datasets are used to generate HDL code. Although these training datasets are limited and do not yet encompass all multi-modal data, significant progress has been made.
    Agents are employed to optimize models for HDL code generation, though their applications remain somewhat narrow and lack comprehensive coverage.
    
    \item \textbf{Level 3: MLM-governed Submodel}.
    In this stage, the MLM-governed submodel of LPCM leverages fully trained Multimodal Large Models (MLMs) to autonomously generate HDL code during the chip design process. These MLMs operate independently, invoking external tools without human intervention.
\end{itemize}

\begin{table}[htbp]
    \centering
    \caption{Comparison of Different Levels of HDL Generation Systems}
    \label{tab:hdl_levels}
    \begin{tabular}{|l|p{0.45\textwidth}|p{0.45\textwidth}|}
        \hline
        \textbf{Level} & \textbf{Does the Model Require Training?} & \textbf{How EDA Tools Are Used?} \\
        \hline
        \textbf{Level 1} & No training: General-purpose LLMs generate HDL code directly. & Passive verification: EDA tools verify HDL code without participating in generation. \\
        \hline
        \textbf{Level 2} & Fine-tuned: Large models adapt to domain-specific datasets for accurate HDL generation. & Active optimization: EDA tools act as agents, refining HDL code via feedback.\\
        \hline
        \textbf{Level 3} & Full training: Multimodal LLMs (MLMs) are trained end-to-end for autonomous HDL design. & Full integration: EDA tools and MLM co-execute autonomous verification and optimization.\\
        \hline
    \end{tabular}
\end{table}

\subsubsection{Related work}
Direct Prompting and Feedback-Driven Optimization. Prior research has explored the direct use of LLMs to generate HDL code. For instance, Chip-chat \cite{Chip-Chat23} designed an 8-bit accumulator-based microprocessor using commercial LLMs. Other studies \cite{AutoChip23,DeLorenzo24} investigate prompt optimization through feedback from EDA tools. AutoChip \cite{AutoChip23} leverages error reports from compilers and simulators to help LLMs correct faulty code, while MEMV \cite{DeLorenzo24} employs a Monte Carlo tree-search algorithm to improve the correctness and PPA efficiency of generated code based on feedback from compilers and synthesis tools. OriGen \cite{OriGen24} uses a self-reflection mechanism guided by compiler feedback to fix syntactic errors in the generated code.

Fine-Tuning Strategies. VeriGen \cite{VeriGen23} performs continual pre-training by predicting the next token on corpora from open-source code and textbooks. VerilogEval \cite{VerilogEval23} applies supervised fine-tuning (SFT) using synthetic instruction-code pairs and releases an open-source evaluation dataset containing 156 questions with golden solutions. MEV-LLM \cite{MEV-LLM24} fine-tunes LLMs for hardware designs of varying complexity and integrates them into a unified framework. ChipNeMo \cite{ChipNeMo24} customizes LLMs for applications, including chatbots, EDA script generation, and bug summarization. RTLCoder \cite{RTLCoder23} introduces a fine-tuning algorithm that evaluates code quality on candidate samples generated by pre-trained LLMs. BetterV \cite{BetterV24} optimizes Verilog code generation using generative discriminators, while CodeV \cite{CodeV24} constructs a fine-tuning dataset by generating multi-level Verilog code summaries with LLMs.

Agent-based approaches improve HDL generation by integrating simulation feedback throughout the code generation pipeline, covering planning, verification, and iterative refinement. RTLLM \cite{RTLLM24} evaluates syntax correctness, functional accuracy, and design quality, while ITERTL \cite{ITERTL24} iteratively optimizes training data using RTL tool feedback. Multi-agent systems \cite{AIvril24, VerilogCoder25, MAGE24} enhance reliability through specialized agents, such as those handling RTL coding, testbench generation, validation, and debugging, collaborating in a recursive framework to produce optimized implementations.

\subsection{Proposed MLM-governed HDL module design}
The MLM-HDL model is trained using chip design data generated by upstream modules. The training data consists of natural language descriptions, circuit diagrams, tables, and C code, paired with the corresponding HDL code. The architecture simulator provides supervisory signals to fine-tune the multimodal model, enabling it to incorporate the physical meaning of chip designs.

During inference, the model accepts design parameters from the DSE Agent and generates optimized HDL code. This output undergoes verification through EDA tools, with simulation-validated feedback enabling continuous design refinement. The system achieves complete automation from high-level specifications to production-ready HDL implementation.

\begin{figure}
    \centering
    \includegraphics[width=1\linewidth]{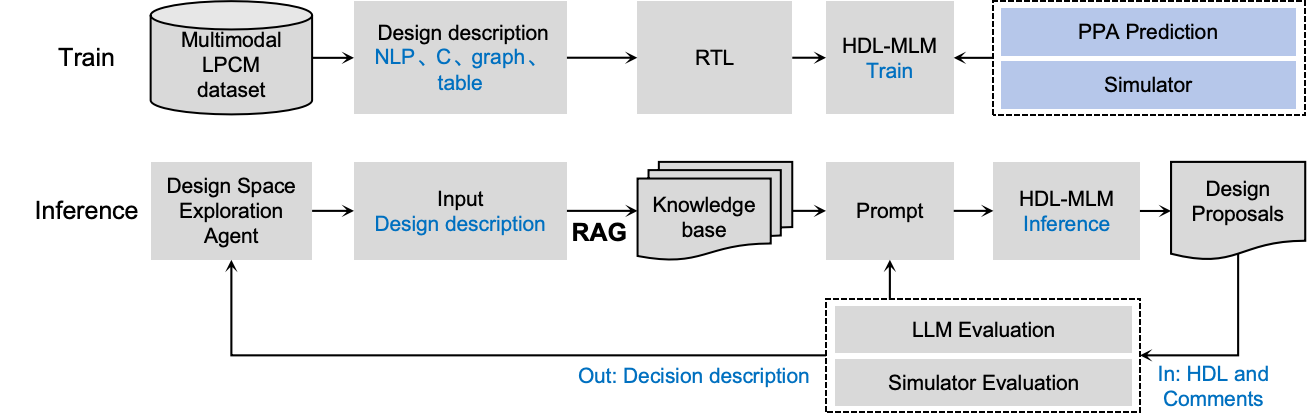}
    \caption{HDL Level 3 workflow: MLM-Governed}
    \label{fig:enter-label}
\end{figure}

Key capabilities include: 

\begin{itemize}
    \item Automated Design Process. The system achieves a fully autonomous design loop where large models can generate complete HDL code from high-level descriptions including functional requirements and performance metrics. This enables end-to-end implementation from initial concept to final physical design without human intervention. The models perform cross-layer optimization by integrating simulation tools, verification platforms, and synthesis tools to ensure the generated code meets specific design constraints such as power consumption, performance, and area requirements. Through hardware-software co-design, they maintain consistency across design layers while achieving overall performance optimization.

    \item Efficient Quality Assurance Based on RTL-Level PPA Prediction. We propose a multimodal PPA prediction model for RTL-stage design, integrating both LLMs and graph neural networks (GNNs). First, we leverage fine-tuned LLMs to encode high-level functional and structural information directly from RTL code, focusing in particular on register and critical logic descriptions. Meanwhile, we map the synthesized netlist into a standard-cell graph, allowing a GNN to capture granular structural and timing dependencies. Through methods such as adaptive aggregation and two-phase propagation, the GNN efficiently models local circuit behaviors, while global functionality insights come from the LLM, achieving more accurate PPA predictions even as circuit size grows.
    On top of this multimodal fusion, our training framework adopts knowledge distillation from a layout-aware “teacher” model to guide the RTL-level “student” model toward near sign-off precision on key metrics like arrival time and power. By aligning node-, subgraph-, and global-level features, the student progressively assimilates timing-critical and layout-specific knowledge. This strategy not only narrows the accuracy gap between RTL and post-layout predictions but also preserves efficiency through a lightweight GNN-based student, making it highly practical for early-stage design exploration.
    
    \item Precision Control and Optimization. A feedback-based learning mechanism enables continuous improvement by incorporating EDA toolchain feedback. When initial designs fail to meet PPA standards, the models dynamically adjust parameters or architectural choices to regenerate optimized HDL code. The system also generates innovative design proposals by incorporating cutting-edge research, such as creating HDL implementations for emerging memory technologies or quantum computing architectures.

    \item Efficient Collaboration Under Human Guidance. The process maintains goal-oriented design where human designers set high-level objectives like "design a low-power IoT processor" while the model handles implementation details. This approach supports rapid iteration and prototype development, allowing designers to quickly test concepts and refine versions, significantly accelerating time-to-market while enhancing design quality and innovation potential.

\end{itemize}

In conclusion, HDL-MLMs are transforming HDL generation and optimization through domain expertise integration, workflow automation, and feedback-driven learning, revolutionizing traditional design methodologies.

%% file: submit_version/10_sam.tex
\section{Put Them All Together}

\begin{figure}
\centering
\includegraphics[width=0.99\linewidth]{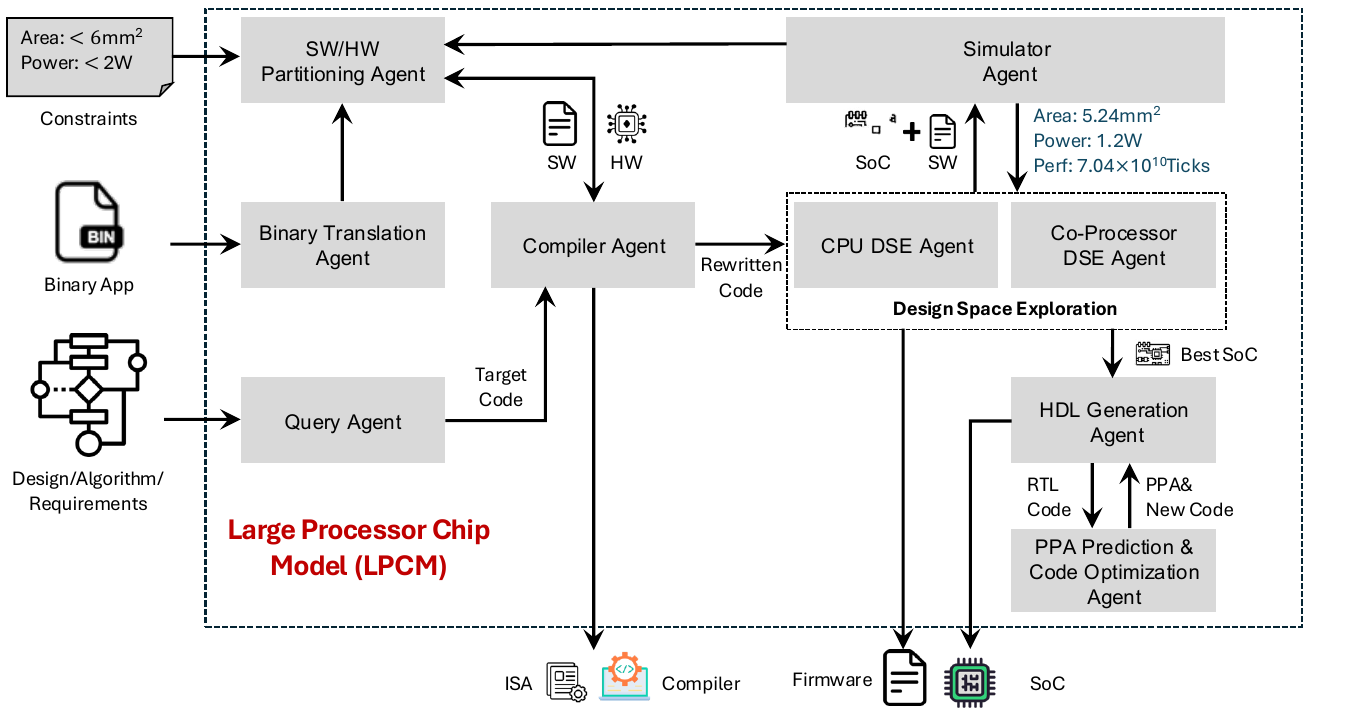}
\caption{The end-to-end LPCM flow}
\label{fig:ptat:architecture}
\end{figure}

As shown in \autoref{fig:ptat:architecture}, the framework accepts two input types: binary applications or high-level design specifications, both of which generate an intermediate target code representation.

The SW/HW Partitioning Agent analyzes this representation to determine optimal hardware-software boundaries while respecting the area and power constraints. This agent works alongside the Compiler Agent, which generates appropriate compiler implementations and processes code according to the specified ISA.
The rewritten code then enters the design space exploration phase conducted by parallel CPU and Co-Processor DSE Agents. These agents work in tandem with the Simulator Agent, continuously refining architectural proposals through iterative evaluation.

Once an optimal SoC design emerges, the HDL Generation Agent produces RTL code, which undergoes further refinement by the PPA Prediction \& Code Optimization Agent. The framework ultimately delivers four key outputs: ISA specifications, compiler implementations, firmware, and the physical SoC.

\subsection{Gradual LLM Integration}
The LPCM framework implements a graduated approach to integrating LLMs into the system architecture design process, progressing through three distinct levels of automation and integration. Each level represents an advancement in the role, capabilities, and autonomy of LLMs within the design workflow.

\begin{itemize}
    \item \textbf{Level 1: Human-Centric with LLM Assistance.} At this level, human designers remain the primary decision-makers, with LLMs functioning as assistive tools. Experts define design requirements, make architectural decisions, and execute core design tasks, while LLMs provide supplementary support such as code generation, documentation assistance, and reference information retrieval. The workflow begins with human designers establishing clear specifications and constraints. LLMs then assist by generating code snippets, suggesting optimization approaches, or retrieving relevant examples from prior designs. Human experts carefully review, modify, and integrate all LLM-generated content into the overall design. Verification and validation processes remain entirely under human control, with LLMs contributing only to specific subtasks within well-defined boundaries. LLMs focus on enhancing designer productivity rather than making autonomous decisions.
    
    \item \textbf{Level 2: Agent-Orchestrated with Multi-submodule Cooperation.} This level represents a significant advancement in automation, where LLM-based agents coordinate activities across multiple design modules with reduced human intervention. These agents possess greater domain-specific knowledge and can handle routine decision-making tasks autonomously while still operating within frameworks established by human designers. Each module incorporates specialized agent capabilities tailored to its domain. For example, in the Compiler module, agents can autonomously perform code analysis, optimization selection, and transformation application. In the Hardware/Software Partitioning module, agents can identify code segments amenable to hardware acceleration and propose partitioning strategies. These module-specific agents coordinate through standardized interfaces, exchanging structured information and collaborating to solve cross-domain challenges. The interaction becomes bidirectional—LLMs analyze performance feedback and propose design adjustments, while experts refine constraints or objectives as needed.
    
    \item \textbf{Level 3: LLM-Governed Autonomous Generation.} This level represents a transformative shift where LLMs assume primary control over the entire design process. Humans provide only high-level objectives and constraints, while LLMs autonomously execute the complete design workflow from requirements analysis through implementation. The LLM-governed system exhibits advanced reasoning capabilities, domain expertise across the entire system stack, and the ability to make sophisticated trade-offs between competing design objectives. The workflow begins with minimal human input, typically a concise description of functional requirements and design constraints. The system then autonomously decomposes these requirements into specific tasks, coordinates execution across modules, evaluates multiple design alternatives, and progressively refines solutions based on performance feedback. It can identify and resolve conflicts between requirements, explore unconventional design approaches, and provide detailed explanations for its design decisions, achieving a high degree of autonomy in hardware design.
\end{itemize}

\subsection{Case Study: 3D Gaussian Splatting}
This section demonstrates the effectiveness of LPCM through a practical application to 3D Gaussian Splatting (3DGS), an emerging rendering technique that has garnered significant attention for its ability to deliver high-quality visual experiences with improved performance. 3DGS represents scenes using a collection of 3D Gaussian primitives, providing a balance between quality and rendering speed that is crucial for applications in virtual reality, scientific visualization, and computer graphics.

The workflow begins with two parallel paths of input processing. While the Binary Translation Agent can process executable applications, in this case study, we focus on the path through the Query Agent. The Query Agent conducts a comprehensive search across multiple GitHub repositories to identify various implementations of the 3DGS rendering algorithm. During this process, the agent evaluates multiple candidate codebases, comparing their quality metrics such as implementation completeness, code efficiency, and documentation quality. After thorough analysis and comparison, the agent selected the implementation from the repository cited in \cite{githubGitHubMmtatdiffgaussianrasterization} as the optimal choice due to its alignment with the design objectives. This particular implementation stood out for its comprehensive feature set and platform versatility, offering a complete implementation that could run efficiently on CPU architecture rather than being highly platform-dependent.  %This selected Target Code serves as the foundation for understanding the algorithm's structure, computational patterns, and potential bottlenecks when executed on general-purpose hardware.

Simultaneously, the framework receives design constraints specifying requirements of area less than 6mm² and power consumption under 2W, which directly inform the SW/HW Partitioning Agent's decision-making process.

The Target Code from the Query Agent is then passed to the Compiler Agent, which compiles it for the target CPU platform, generating an executable binary with comprehensive debug information. This debug-enabled binary preserves the algorithmic essence while enabling detailed runtime analysis. Then the SW/HW Partitioning Agent uses this instrumented executable to identify acceleration candidates, recognizing pixel coordinate transformations, conic term calculations, and power sum computations as prime hardware acceleration targets, while maintaining control logic in software. The resulting design specifies clear interfaces between software and hardware components.
\begin{figure}
\centering
\includegraphics[width=0.99\linewidth]{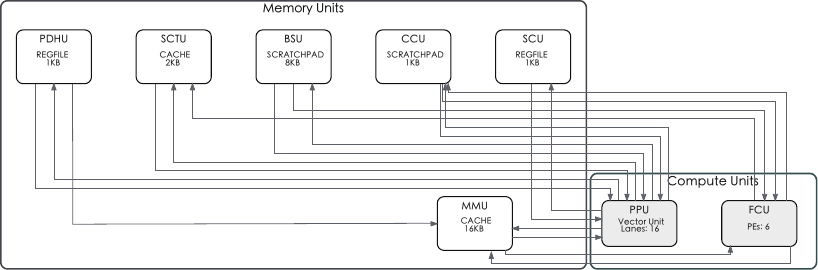}
\caption{The 3DGS accelerator designed by the LPCM flow}
\label{fig:ptat:case_study}
\end{figure}

Based on this partitioning, the Compiler Agent generates optimized code for both software execution and hardware implementation. This involves producing rewritten code that incorporates extended instructions to access hardware-accelerated components. The compiler specifically creates custom instruction extensions that allow the CPU to efficiently communicate with and offload computations to the specialized co-processor units. In addition, this stage also outputs detailed specifications for these extended instructions that define their functionality and interface, and a modified compiler toolchain that supports the new extension to the instruction set.

The Design Space Exploration Agent, comprising the Co-Processor DSE Agent and CPU DSE Agent, evaluates various architectural configurations to optimize performance under the given constraints. For 3DGS, this exploration focuses on finding the optimal balance between computational units, memory hierarchy, and the interconnect structure. The exploration process evaluates different configurations of CPU and co-processor choices and their parameters, and memory sizes to maximize rendering throughput while meeting the area and power constraints.

Following the identification of the optimal architecture, the HDL Generation Agent translates the architectural specification into RTL code that describes the hardware implementation. For the 3DGS accelerator, this includes generating descriptions for specialized units such as the Pixel Processing Unit with its 16-lane vector unit, the Feature Computation Unit with its 6 processing elements, and memory management units with their respective storage capacities, as shown in \autoref{fig:ptat:case_study}.

Finally, the PPA Prediction \& Code Optimization Agent refines the implementation to ensure it meets the target constraints. For our 3DGS case, the final design achieves an area of 5.24mm² and power consumption of 1.2W, comfortably within the specified constraints. Performance analysis shows a 1.41× speedup over a high-performance NVIDIA A100 GPU while requiring only a fraction of the chip area.

\subsection{Challenges and Future Planning}
As we advance our Large Processor Chip Model toward higher levels of automation and integration, several significant challenges must be addressed. This section outlines these challenges and our strategies for overcoming them.

\subsubsection{Multi-modal Knowledge Integration and Transfer}
A significant challenge lies in integrating and transferring knowledge across diverse modalities and domains within the LPCM framework. Each module operates on different data representations—including source code, compiler intermediate representations, task graphs, architectural specifications, and hardware description languages—making knowledge sharing inherently difficult. Furthermore, optimization techniques that prove effective in one domain may not directly translate to others without substantial adaptation.

To address this challenge, we are developing advanced multi-modal learning frameworks capable of extracting, representing, and transferring knowledge across domains. Our approach leverages techniques from transfer learning and multi-task learning to identify generalizable patterns and principles that apply across abstraction levels. By training models on paired examples from different domains, such as software implementations and corresponding hardware accelerators.

We are also establishing comprehensive knowledge repositories that capture design patterns, optimization strategies, and architectural templates across domains. These repositories employ structured knowledge representations that facilitate retrieval and application in new design contexts. By explicitly modeling the relationships between patterns in different domains, we can more effectively transfer successful approaches across abstraction boundaries. This knowledge-sharing infrastructure is complemented by continual learning mechanisms that progressively enhance the system's capabilities based on design experience, enabling it to recognize increasingly complex cross-domain opportunities.

\subsubsection{Verification and Trustworthiness of Autonomous Design Decisions}
As the LPCM framework progresses toward higher levels of automation, ensuring the verification and trustworthiness of autonomously generated designs becomes increasingly critical. Traditional verification methodologies that rely heavily on human oversight become impractical as the system assumes greater responsibility for design decisions. Establishing confidence in the correctness, optimality, and robustness of automatically generated architectures presents significant technical and methodological challenges.

Our approach to addressing this challenge incorporates multiple complementary strategies. First, we are integrating formal verification techniques throughout the design flow. These techniques provide mathematical guarantees about specific properties of the generated designs, such as functional correctness, timing compliance, and security characteristics. By formally verifying critical aspects of the design, we can provide stronger assurances than possible with traditional simulation-based approaches. We are particularly focused on compositional verification methods that can scale to complex system architectures by verifying components individually and their interactions systematically.

Second, we are establishing comprehensive validation frameworks that test generated designs against diverse workloads and operating conditions. These frameworks employ systematic testing strategies to explore the design's behavior across nominal and edge cases, identifying potential weaknesses before implementation. By subjecting designs to rigorous validation, we can identify and address potential issues early in the development process, increasing confidence in the final implementation.

Finally, we are developing incremental automation approaches that gradually transfer responsibility from human designers to the autonomous system as confidence in its capabilities increases. This progressive autonomy enables the system to establish a track record of successful designs in limited contexts before tackling more complex challenges. By carefully managing this transition, we can build trust in the system's capabilities while minimizing the risks associated with autonomous design decisions.
\subsubsection{Cross-Layer Optimization in Heterogeneous Design Spaces}
A fundamental challenge in developing an effective LPCM framework lies in achieving true cross-layer optimization across heterogeneous design spaces. Traditional design methodologies typically optimize each architectural layer independently—application, algorithm, compiler, instruction set, microarchitecture, and circuit implementation, leading to suboptimal global solutions. When design decisions at one layer constrain options at another without coordination, the resulting architectures often exhibit inefficiencies that could be avoided through holistic optimization. This challenge is exacerbated in domain-specific architectures, where the interdependencies between software and hardware design choices become increasingly complex and the potential performance gains from cross-layer optimization grow substantially.

The difficulty stems from several factors: first, the design spaces at different layers are inherently heterogeneous, with distinct representations and optimization metrics that complicate joint exploration. Second, the causal relationships between design decisions across layers are often complex and non-obvious, making it difficult to predict how changes at one layer will propagate through the system. Third, the sheer dimensionality of the combined design space makes exhaustive exploration infeasible, necessitating intelligent navigation strategies that can identify promising cross-layer optimization opportunities.

To address this challenge, we are pursuing several innovative approaches. One promising direction involves reinforcement learning finetuning techniques for cross-layer optimization. By formulating the design process as a sequential decision-making problem, we can train RL agents to make coordinated design choices across multiple layers simultaneously. These agents learn to navigate the complex cross-layer design space through experience, discovering optimization strategies that may not be apparent through conventional methods. The RL framework enables explicit modeling of the long-term performance implications of design decisions, allowing the system to make early-stage choices that facilitate more effective optimization at later stages.

Another innovative approach we are exploring is Critique Finetuning for knowledge alignment across architectural domains. This method leverages expert feedback to critically evaluate design decisions and their cross-layer implications, progressively refining the system's understanding of effective design patterns. 

%% file: submit_version/11_conclusion.tex
\section{Conclusion}
Research in computer system architecture has evolved significantly, spanning from quantum computing to large-scale data centers, characterized by an unprecedented rate of development, broad impact, and deep influence. Nevertheless, current approaches in this field continue to depend heavily on the conventional process of "research design-experimental verification-data analysis," which presents challenges such as inefficiency, variable quality, and difficulties in replicability. The rise of large language models (LLMs), noted for their sophisticated learning, reasoning, and planning capabilities, offers exciting opportunities for reshaping the paradigm of computer system architecture research. In this context, this paper presents a Large Processor Chip Model (LPCM). It organizes its development into three levels and explores various dimensions, including task complexity, model processing capabilities, and trends in technological evolution. This paper further explores the specific realization paths and technical details related to six key aspects: compiler design, binary translation, simulator design, hardware and software partitioning, design space exploration, and RTL design and generation, across various levels. Additionally, this paper utilizes 3D Gaussian Splatting (3DGS) as a representative workload to examine the LPCM design methodology and operational flow at Level 1, emphasizing the benefits of LPCM in improving design efficiency and quality. Finally, this paper provides an in-depth analysis of the challenges and potential solutions related to the development of LPCM. In conclusion, LPCM is poised to transform the current methodologies of computer system architecture design, steering the information infrastructure towards greater intelligence and automation, while paving the way for new advancements in information technology.

%% file: submit_main.bbl
\begin{thebibliography}{10}
\providecommand{\url}[1]{\texttt{#1}}
\providecommand{\urlprefix}{URL }
\providecommand{\bibinfo}[2]{#2}

\bibitem{hennessy2019new}
\bibinfo{author}{Hennessy J~L}, \bibinfo{author}{Patterson D~A}.
\newblock \bibinfo{title}{A new golden age for computer architecture}.
\newblock \bibinfo{journal}{Communications of the ACM}, \bibinfo{year}{2019},
  \bibinfo{volume}{62}: \bibinfo{pages}{48--60}

\bibitem{ABC}
\bibinfo{author}{{Berkeley Logic Synthesis and Verification Group}}.
\newblock \bibinfo{title}{Abc: A system for sequential logic synthesis and
  formal verification}.
\newblock \urlprefix\url{https://people.eecs.berkeley.edu/~alanmi/abc/}

\bibitem{EdwardsOpenCircuit}
\bibinfo{author}{Edwards T}.
\newblock \bibinfo{title}{Open circuit design}.
\newblock \urlprefix\url{http://opencircuitdesign.com/}

\bibitem{OpenSTA}
\bibinfo{author}{{The OpenROAD Project}}.
\newblock \bibinfo{title}{Opensta: A static timing analysis tool}.
\newblock \urlprefix\url{https://github.com/The-OpenROAD-Project/OpenSTA}

\bibitem{IcarusVerilog}
\bibinfo{author}{{Icarus Verilog Development Team}}.
\newblock \bibinfo{title}{Icarus verilog}.
\newblock \urlprefix\url{https://steveicarus.github.io/iverilog/}

\bibitem{Yosys}
\bibinfo{author}{Wolf C}, \bibinfo{author}{Glaser J}.
\newblock \bibinfo{title}{Yosys - a free verilog synthesis suite}.
\newblock \urlprefix\url{https://github.com/YosysHQ/yosys}

\bibitem{schafer2009adaptive}
\bibinfo{author}{Schafer B~C}, \bibinfo{author}{Takenaka T},
  \bibinfo{author}{Wakabayashi K}.
\newblock \bibinfo{title}{Adaptive simulated annealer for high level synthesis
  design space exploration}.
\newblock In: \bibinfo{booktitle}{2009 International Symposium on VLSI Design,
  Automation and Test}, \bibinfo{year}{2009}.
\newblock \bibinfo{pages}{106--109}

\bibitem{mahapatra2014machine}
\bibinfo{author}{Mahapatra A}, \bibinfo{author}{Schafer B~C}.
\newblock \bibinfo{title}{Machine-learning based simulated annealer method for
  high level synthesis design space exploration}.
\newblock In: \bibinfo{booktitle}{Proceedings of the 2014 Electronic System
  Level Synthesis Conference (ESLsyn)}, \bibinfo{year}{2014}.
\newblock \bibinfo{pages}{1--6}

\bibitem{brayton1984logic}
\bibinfo{author}{Brayton R~K}, \bibinfo{author}{Hachtel G~D},
  \bibinfo{author}{McMullen C}, et~al.
\newblock \bibinfo{title}{Logic minimization algorithms for VLSI synthesis},
  volume~\bibinfo{volume}{2}.
\newblock \bibinfo{year}{1984}, \bibinfo{year}{1984}

\bibitem{keutzer1987dagon}
\bibinfo{author}{Keutzer K}.
\newblock \bibinfo{title}{Dagon: Technology binding and local optimization by
  dag matching}.
\newblock In: \bibinfo{booktitle}{Proceedings of the 24th ACM/IEEE Design
  Automation Conference}, \bibinfo{year}{1987}.
\newblock \bibinfo{pages}{341--347}

\bibitem{grosnit2022boils}
\bibinfo{author}{Grosnit A}, \bibinfo{author}{Malherbe C},
  \bibinfo{author}{Tutunov R}, et~al.
\newblock \bibinfo{title}{Boils: Bayesian optimisation for logic synthesis}.
\newblock In: \bibinfo{booktitle}{2022 Design, Automation \& Test in Europe
  Conference \& Exhibition (DATE)}, \bibinfo{year}{2022}.
\newblock \bibinfo{pages}{1193--1196}

\bibitem{cheng2018replace}
\bibinfo{author}{Cheng C~K}, \bibinfo{author}{Kahng A~B}, \bibinfo{author}{Kang
  I}, et~al.
\newblock \bibinfo{title}{Replace: Advancing solution quality and routability
  validation in global placement}.
\newblock \bibinfo{journal}{IEEE Transactions on Computer-Aided Design of
  Integrated Circuits and Systems}, \bibinfo{year}{2018}, \bibinfo{volume}{38}:
  \bibinfo{pages}{1717--1730}

\bibitem{singh1988timing}
\bibinfo{author}{Singh K~J}, \bibinfo{author}{Wang A~R},
  \bibinfo{author}{Brayton R~K}, et~al.
\newblock \bibinfo{title}{Timing optimization of combinational logic}.
\newblock In: \bibinfo{booktitle}{1988 IEEE International Conference on
  Computer-Aided Design}, \bibinfo{year}{1988}.
\newblock \bibinfo{pages}{282--283}

\bibitem{ajayi2019toward}
\bibinfo{author}{Ajayi T}, \bibinfo{author}{Chhabria V~A},
  \bibinfo{author}{Foga{\c{c}}a M}, et~al.
\newblock \bibinfo{title}{Toward an open-source digital flow: First learnings
  from the openroad project}.
\newblock In: \bibinfo{booktitle}{Proceedings of the 56th Annual Design
  Automation Conference 2019}, \bibinfo{year}{2019}.
\newblock \bibinfo{pages}{1--4}

\bibitem{carrion2012machine}
\bibinfo{author}{Carrion~Schafer B}, \bibinfo{author}{Wakabayashi K}.
\newblock \bibinfo{title}{Machine learning predictive modelling high-level
  synthesis design space exploration}.
\newblock \bibinfo{journal}{IET computers \& digital techniques},
  \bibinfo{year}{2012}, \bibinfo{volume}{6}: \bibinfo{pages}{153--159}

\bibitem{zuluaga2012smart}
\bibinfo{author}{Zuluaga M}, \bibinfo{author}{Krause A},
  \bibinfo{author}{Milder P}, et~al.
\newblock \bibinfo{title}{" smart" design space sampling to predict
  pareto-optimal solutions}.
\newblock In: \bibinfo{booktitle}{Proceedings of the 13th ACM SIGPLAN/SIGBED
  International Conference on Languages, Compilers, Tools and Theory for
  Embedded Systems}, \bibinfo{year}{2012}.
\newblock \bibinfo{pages}{119--128}

\bibitem{ferianc2020improving}
\bibinfo{author}{Ferianc M}, \bibinfo{author}{Fan H}, \bibinfo{author}{Chu
  R~S}, et~al.
\newblock \bibinfo{title}{Improving performance estimation for fpga-based
  accelerators for convolutional neural networks}.
\newblock In: \bibinfo{booktitle}{Applied Reconfigurable Computing.
  Architectures, Tools, and Applications: 16th International Symposium, ARC
  2020, Toledo, Spain, April 1--3, 2020, Proceedings 16}, \bibinfo{year}{2020}.
\newblock \bibinfo{pages}{3--13}

\bibitem{mirhoseini2020chip}
\bibinfo{author}{Mirhoseini A}, \bibinfo{author}{Goldie A},
  \bibinfo{author}{Yazgan M}, et~al.
\newblock \bibinfo{title}{Chip placement with deep reinforcement learning}.
\newblock \bibinfo{journal}{arXiv preprint arXiv:2004.10746},
  \bibinfo{year}{2020}

\bibitem{he2020circuit}
\bibinfo{author}{He Y}, \bibinfo{author}{Bao F~S}.
\newblock \bibinfo{title}{Circuit routing using monte carlo tree search and
  deep neural networks}.
\newblock \bibinfo{journal}{arXiv preprint arXiv:2006.13607},
  \bibinfo{year}{2020}

\bibitem{alawieh2020high}
\bibinfo{author}{Alawieh M~B}, \bibinfo{author}{Li W}, \bibinfo{author}{Lin Y},
  et~al.
\newblock \bibinfo{title}{High-definition routing congestion prediction for
  large-scale {FPGAs}}.
\newblock In: \bibinfo{booktitle}{Proceedings of the Asia and South Pacific
  Design Automation Conference}, \bibinfo{year}{2020}.
\newblock \bibinfo{pages}{26--31}

\bibitem{ChipNeMo24}
\bibinfo{author}{Liu M}, \bibinfo{author}{Ene T}, \bibinfo{author}{Kirby R},
  et~al.
\newblock \bibinfo{title}{Chipnemo: Domain-adapted llms for chip design}.
\newblock \bibinfo{journal}{CoRR}, \bibinfo{year}{2023},
  \bibinfo{volume}{abs/2311.00176}.
\newblock \urlprefix\url{https://doi.org/10.48550/arXiv.2311.00176}

\bibitem{RTLCoder23}
\bibinfo{author}{Liu S}, \bibinfo{author}{Fang W}, \bibinfo{author}{Lu Y},
  et~al.
\newblock \bibinfo{title}{Rtlcoder: Outperforming {GPT-3.5} in design {RTL}
  generation with our open-source dataset and lightweight solution}.
\newblock \bibinfo{journal}{CoRR}, \bibinfo{year}{2023},
  \bibinfo{volume}{abs/2312.08617}.
\newblock \urlprefix\url{https://doi.org/10.48550/arXiv.2312.08617}

\bibitem{BetterV24}
\bibinfo{author}{Pei Z}, \bibinfo{author}{Zhen H}, \bibinfo{author}{Yuan M},
  et~al.
\newblock \bibinfo{title}{Betterv: Controlled verilog generation with
  discriminative guidance}.
\newblock In: \bibinfo{booktitle}{Forty-first International Conference on
  Machine Learning, {ICML} 2024, Vienna, Austria, July 21-27, 2024},
  \bibinfo{year}{2024}.
\newblock \urlprefix\url{https://openreview.net/forum?id=jKnW7r7de1}

\bibitem{wu2024chateda}
\bibinfo{author}{Wu H}, \bibinfo{author}{He Z}, \bibinfo{author}{Zhang X},
  et~al.
\newblock \bibinfo{title}{Chateda: A large language model powered autonomous
  agent for eda}.
\newblock \bibinfo{journal}{IEEE Transactions on Computer-Aided Design of
  Integrated Circuits and Systems}, \bibinfo{year}{2024}

\bibitem{ChipGPT23}
\bibinfo{author}{Chang K}, \bibinfo{author}{Wang Y}, \bibinfo{author}{Ren H},
  et~al.
\newblock \bibinfo{title}{Chipgpt: How far are we from natural language
  hardware design}.
\newblock \bibinfo{journal}{CoRR}, \bibinfo{year}{2023},
  \bibinfo{volume}{abs/2305.14019}.
\newblock \urlprefix\url{https://doi.org/10.48550/arXiv.2305.14019}

\bibitem{fang2024assertllm}
\bibinfo{author}{Fang W}, \bibinfo{author}{Li M}, \bibinfo{author}{Li M},
  et~al.
\newblock \bibinfo{title}{Assertllm: Generating and evaluating hardware
  verification assertions from design specifications via multi-llms}.
\newblock \bibinfo{journal}{arXiv preprint arXiv:2402.00386},
  \bibinfo{year}{2024}

\bibitem{tsai2024rtlfixer}
\bibinfo{author}{Tsai Y}, \bibinfo{author}{Liu M}, \bibinfo{author}{Ren H}.
\newblock \bibinfo{title}{Rtlfixer: Automatically fixing rtl syntax errors with
  large language model}.
\newblock In: \bibinfo{booktitle}{Proceedings of the 61st ACM/IEEE Design
  Automation Conference}, \bibinfo{year}{2024}.
\newblock \bibinfo{pages}{1--6}

\bibitem{smt-ASE15}
\bibinfo{author}{Nguyen A~T}, \bibinfo{author}{Nguyen T~T},
  \bibinfo{author}{Nguyen T~N}.
\newblock \bibinfo{title}{Divide-and-conquer approach for multi-phase
  statistical migration for source code (t)}.
\newblock In: \bibinfo{booktitle}{2015 30th IEEE/ACM International Conference
  on Automated Software Engineering (ASE)}, \bibinfo{year}{2015}.
\newblock \bibinfo{pages}{585--596}

\bibitem{tree2tree-NIPS18}
\bibinfo{author}{Chen X}, \bibinfo{author}{Liu C}, \bibinfo{author}{Song D}.
\newblock \bibinfo{title}{Tree-to-tree neural networks for program
  translation}.
\newblock In: \bibinfo{booktitle}{Proceedings of the 32nd International
  Conference on Neural Information Processing Systems}, \bibinfo{year}{2018}.
\newblock \bibinfo{pages}{2552–2562}

\bibitem{feng2020codebert}
\bibinfo{author}{Feng Z}, \bibinfo{author}{Guo D}, \bibinfo{author}{Tang D},
  et~al.
\newblock \bibinfo{title}{Codebert: A pre-trained model for programming and
  natural languages}, \bibinfo{year}{2020}

\bibitem{wang2021codet5}
\bibinfo{author}{Wang Y}, \bibinfo{author}{Wang W}, \bibinfo{author}{Joty S},
  et~al.
\newblock \bibinfo{title}{Codet5: Identifier-aware unified pre-trained
  encoder-decoder models for code understanding and generation},
  \bibinfo{year}{2021}

\bibitem{zhong2024comback}
\bibinfo{author}{Zhong M}, \bibinfo{author}{Lyu F}, \bibinfo{author}{Wang L},
  et~al.
\newblock \bibinfo{title}{Comback: A versatile dataset for enhancing compiler
  backend development efficiency}.
\newblock In: \bibinfo{booktitle}{The Thirty-eight Conference on Neural
  Information Processing Systems Datasets and Benchmarks Track},
  \bibinfo{year}{2024}

\bibitem{zhang-etal-2024-introducing}
\bibinfo{author}{Zhang S}, \bibinfo{author}{Zhao J}, \bibinfo{author}{Xia C},
  et~al.
\newblock \bibinfo{title}{Introducing compiler semantics into large language
  models as programming language translators: A case study of {C} to x86
  assembly}.
\newblock \bibinfo{year}{2024}: \bibinfo{pages}{996--1011}.
\newblock \urlprefix\url{https://aclanthology.org/2024.findings-emnlp.55/}

\bibitem{chen2021vegen}
\bibinfo{author}{Chen Y}, \bibinfo{author}{Mendis C}, \bibinfo{author}{Carbin
  M}, et~al.
\newblock \bibinfo{title}{Vegen: a vectorizer generator for simd and beyond}.
\newblock In: \bibinfo{booktitle}{Proceedings of the 26th ACM International
  Conference on Architectural Support for Programming Languages and Operating
  Systems}, \bibinfo{year}{2021}.
\newblock \bibinfo{pages}{902--914}

\bibitem{armengol2021learning}
\bibinfo{author}{Armengol-Estap{\'e} J}, \bibinfo{author}{O'Boyle M~F}.
\newblock \bibinfo{title}{Learning c to x86 translation: An experiment in
  neural compilation}.
\newblock \bibinfo{journal}{arXiv preprint arXiv:2108.07639},
  \bibinfo{year}{2021}

\bibitem{c2llvm-HPEC22}
\bibinfo{author}{Guo Z~C}, \bibinfo{author}{Moses W~S}.
\newblock \bibinfo{title}{Enabling transformers to understand low-level
  programs}.
\newblock In: \bibinfo{booktitle}{2022 IEEE High Performance Extreme Computing
  Conference (HPEC)}, \bibinfo{year}{2022}.
\newblock \bibinfo{pages}{1--9}

\bibitem{bolt2023}
\bibinfo{author}{{facebookarchive}}.
\newblock \bibinfo{title}{Bolt: Binary optimization and layout tool}.
\newblock
  \bibinfo{howpublished}{\url{https://github.com/facebookarchive/BOLT}},
  \bibinfo{year}{2023}.
\newblock \bibinfo{note}{Archived on Jul 1, 2023. Part of the LLVM project.}

\bibitem{panchenko2021lightning}
\bibinfo{author}{Panchenko M}, \bibinfo{author}{Auler R},
  \bibinfo{author}{Sakka L}, et~al.
\newblock \bibinfo{title}{Lightning bolt: powerful, fast, and scalable binary
  optimization}.
\newblock In: \bibinfo{booktitle}{Proceedings of the 30th ACM SIGPLAN
  International Conference on Compiler Construction}, \bibinfo{year}{2021}.
\newblock \bibinfo{pages}{119--130}

\bibitem{wong2023refining}
\bibinfo{author}{Wong W~K}, \bibinfo{author}{Wang H}, \bibinfo{author}{Li Z},
  et~al.
\newblock \bibinfo{title}{Refining decompiled c code with large language
  models}.
\newblock \bibinfo{journal}{arXiv preprint arXiv:2310.06530},
  \bibinfo{year}{2023}

\bibitem{armengol2024forklift}
\bibinfo{author}{Armengol-Estap{\'e} J}, \bibinfo{author}{Rocha R~C},
  \bibinfo{author}{Woodruff J}, et~al.
\newblock \bibinfo{title}{Forklift: An extensible neural lifter}.
\newblock \bibinfo{journal}{arXiv preprint arXiv:2404.16041},
  \bibinfo{year}{2024}

\bibitem{tan2024llm4decompile}
\bibinfo{author}{Tan H}, \bibinfo{author}{Luo Q}, \bibinfo{author}{Li J},
  et~al.
\newblock \bibinfo{title}{Llm4decompile: Decompiling binary code with large
  language models}.
\newblock \bibinfo{journal}{arXiv preprint arXiv:2403.05286},
  \bibinfo{year}{2024}

\bibitem{gem5}
\bibinfo{author}{Binkert N}, \bibinfo{author}{Beckmann B},
  \bibinfo{author}{Black G}, et~al.
\newblock \bibinfo{title}{The gem5 simulator}.
\newblock \bibinfo{journal}{SIGARCH Comput. Archit. News},
  \bibinfo{year}{2011}, \bibinfo{volume}{39}: \bibinfo{pages}{1–7}.
\newblock \urlprefix\url{https://doi.org/10.1145/2024716.2024718}

\bibitem{gem5v20}
\bibinfo{author}{Lowe-Power J}, \bibinfo{author}{Ahmad A~M},
  \bibinfo{author}{Akram A}, et~al.
\newblock \bibinfo{title}{The gem5 simulator: Version 20.0+},
  \bibinfo{year}{2020}.
\newblock \urlprefix\url{https://arxiv.org/abs/2007.03152}

\bibitem{booksim}
\bibinfo{author}{Jiang N}, \bibinfo{author}{Becker D~U},
  \bibinfo{author}{Michelogiannakis G}, et~al.
\newblock \bibinfo{title}{A detailed and flexible cycle-accurate
  network-on-chip simulator}.
\newblock In: \bibinfo{booktitle}{2013 IEEE International Symposium on
  Performance Analysis of Systems and Software (ISPASS)}, \bibinfo{year}{2013}.
\newblock \bibinfo{pages}{86--96}

\bibitem{qemu}
\bibinfo{author}{Bellard F}.
\newblock \bibinfo{title}{Qemu, a fast and portable dynamic translator}.
\newblock \bibinfo{year}{2005}: \bibinfo{pages}{41}

\bibitem{simulator-dsv3}
\bibinfo{author}{DeepSeek-AI}, \bibinfo{author}{Liu A}, \bibinfo{author}{Feng
  B}, et~al.
\newblock \bibinfo{title}{Deepseek-v3 technical report}, \bibinfo{year}{2025}.
\newblock \urlprefix\url{https://arxiv.org/abs/2412.19437}

\bibitem{simulator-3dgs}
\bibinfo{author}{Kerbl B}, \bibinfo{author}{Kopanas G},
  \bibinfo{author}{Leimk{\"u}hler T}, et~al.
\newblock \bibinfo{title}{3d gaussian splatting for real-time radiance field
  rendering}.
\newblock \bibinfo{journal}{ACM Transactions on Graphics},
  \bibinfo{year}{2023}, \bibinfo{volume}{42}.
\newblock
  \urlprefix\url{https://repo-sam.inria.fr/fungraph/3d-gaussian-splatting/}

\bibitem{gem5-aladdin}
\bibinfo{author}{Shao Y~S}, \bibinfo{author}{Xi S~L},
  \bibinfo{author}{Srinivasan V}, et~al.
\newblock \bibinfo{title}{Co-designing accelerators and soc interfaces using
  gem5-aladdin}.
\newblock In: \bibinfo{booktitle}{2016 49th Annual IEEE/ACM International
  Symposium on Microarchitecture (MICRO)}, \bibinfo{year}{2016}.
\newblock \bibinfo{pages}{1--12}

\bibitem{eles1997system}
\bibinfo{author}{Eles P}, \bibinfo{author}{Peng Z}, \bibinfo{author}{Kuchcinski
  K}, et~al.
\newblock \bibinfo{title}{System level hardware/software partitioning based on
  simulated annealing and tabu search}.
\newblock \bibinfo{journal}{Design automation for embedded systems},
  \bibinfo{year}{1997}, \bibinfo{volume}{2}: \bibinfo{pages}{5--32}

\bibitem{lo1988heuristic}
\bibinfo{author}{Lo V~M}.
\newblock \bibinfo{title}{Heuristic algorithms for task assignment in
  distributed systems}.
\newblock \bibinfo{journal}{IEEE Transactions on computers},
  \bibinfo{year}{1988}, \bibinfo{volume}{37}: \bibinfo{pages}{1384--1397}

\bibitem{holland1992genetic}
\bibinfo{author}{Holland J~H}.
\newblock \bibinfo{title}{Genetic algorithms}.
\newblock \bibinfo{journal}{Scientific american}, \bibinfo{year}{1992},
  \bibinfo{volume}{267}: \bibinfo{pages}{66--73}

\bibitem{zou2004hw}
\bibinfo{author}{Zou Y}, \bibinfo{author}{Zhuang Z}, \bibinfo{author}{Chen H}.
\newblock \bibinfo{title}{Hw-sw partitioning based on genetic algorithm}.
\newblock In: \bibinfo{booktitle}{Proceedings of the 2004 Congress on
  Evolutionary Computation (IEEE Cat. No. 04TH8753)}, \bibinfo{year}{2004},
  volume~\bibinfo{volume}{1}.
\newblock \bibinfo{pages}{628--633}

\bibitem{van1987simulated}
\bibinfo{author}{Van~Laarhoven P~J}, \bibinfo{author}{Aarts E~H},
  \bibinfo{author}{van Laarhoven P~J}, et~al.
\newblock \bibinfo{title}{Simulated annealing}.
\newblock \bibinfo{year}{1987}, \bibinfo{year}{1987}

\bibitem{selman2006hill}
\bibinfo{author}{Selman B}, \bibinfo{author}{Gomes C~P}.
\newblock \bibinfo{title}{Hill-climbing search}.
\newblock \bibinfo{journal}{Encyclopedia of cognitive science},
  \bibinfo{year}{2006}, \bibinfo{volume}{81}: \bibinfo{pages}{10}

\bibitem{NN_06_ASPLOS}
\bibinfo{author}{Ipek E}, \bibinfo{author}{McKee S~A}, \bibinfo{author}{Caruana
  R}, et~al.
\newblock \bibinfo{title}{Efficiently exploring architectural design spaces via
  predictive modeling}.
\newblock In: \bibinfo{editor}{J~P Shen}, \bibinfo{editor}{M~Martonosi}, eds.,
  \bibinfo{booktitle}{Proceedings of the 12th {{International Conference}} on
  {{Architectural Support}} for {{Programming Languages}} and {{Operating
  Systems}}, {{ASPLOS}} 2006, {{San Jose}}, {{CA}}, {{USA}}, {{October}} 21-25,
  2006}, \bibinfo{year}{2006}.
\newblock \bibinfo{pages}{195--206}

\bibitem{ActBoost_DAC_16}
\bibinfo{author}{Li D}, \bibinfo{author}{Yao S}, \bibinfo{author}{Liu Y},
  et~al.
\newblock \bibinfo{title}{Efficient design space exploration via statistical
  sampling and adaboost learning}.
\newblock In: \bibinfo{booktitle}{DAC, Austin, TX, USA, June 5-9, 2016},
  \bibinfo{year}{2016}.
\newblock \bibinfo{pages}{142:1--142:6}

\bibitem{MoDSE_DAC_23}
\bibinfo{author}{Wang D}, \bibinfo{author}{Yan M}, \bibinfo{author}{Liu X},
  et~al.
\newblock \bibinfo{title}{A {{High-accurate Multi-objective Exploration
  Framework}} for {{Design Space}} of {{CPU}}}.
\newblock In: \bibinfo{booktitle}{{{DAC}}'23: 60th {{ACM}}/{{IEEE Design
  Automation Conference}}, {{San Francisco}}, {{California}}, {{USA}}, {{July}}
  9 - 13, 2023}, \bibinfo{year}{2023}

\bibitem{MoDSE_TCAD_23}
\bibinfo{author}{Wang D}, \bibinfo{author}{Yan M}, \bibinfo{author}{Teng Y},
  et~al.
\newblock \bibinfo{title}{{MoDSE}: A high-accurate multiobjective design space
  exploration framework for {CPU} microarchitectures}.
\newblock \bibinfo{volume}{43}: \bibinfo{pages}{1525--1537}

\bibitem{Linear_model_05_HPCA}
\bibinfo{author}{Joseph P~J}, \bibinfo{author}{Vaswani K},
  \bibinfo{author}{Thazhuthaveetil M~J}.
\newblock \bibinfo{title}{Construction and use of linear regression models for
  processor performance analysis}.
\newblock In: \bibinfo{booktitle}{12th {{International Symposium}} on
  {{High-Performance Computer Architecture}}, {{HPCA-12}} 2006, {{Austin}},
  {{Texas}}, {{USA}}, {{February}} 11-15, 2006}, \bibinfo{year}{2006}.
\newblock \bibinfo{pages}{99--108}

\bibitem{Spline_06_ASPLOS}
\bibinfo{author}{Lee B~C}, \bibinfo{author}{Brooks D~M}.
\newblock \bibinfo{title}{Accurate and efficient regression modeling for
  microarchitectural performance and power prediction}.
\newblock In: \bibinfo{editor}{J~P Shen}, \bibinfo{editor}{M~Martonosi}, eds.,
  \bibinfo{booktitle}{Proceedings of the 12th {{International Conference}} on
  {{Architectural Support}} for {{Programming Languages}} and {{Operating
  Systems}}, {{ASPLOS}} 2006, {{San Jose}}, {{CA}}, {{USA}}, {{October}} 21-25,
  2006}, \bibinfo{year}{2006}.
\newblock \bibinfo{pages}{185--194}

\bibitem{ArchRanker_14_ISCA}
\bibinfo{author}{Chen T}, \bibinfo{author}{Guo Q}, \bibinfo{author}{Tang K},
  et~al.
\newblock \bibinfo{title}{{{ArchRanker}}: {{A}} ranking approach to design
  space exploration}.
\newblock In: \bibinfo{booktitle}{{{ACM}}/{{IEEE}} 41st {{International
  Symposium}} on {{Computer Architecture}}, {{ISCA}} 2014, {{Minneapolis}},
  {{MN}}, {{USA}}, {{June}} 14-18, 2014}, \bibinfo{year}{2014}.
\newblock \bibinfo{pages}{85--96}

\bibitem{AttentionDSE}
\bibinfo{author}{Xue R}, \bibinfo{author}{Wu H}, \bibinfo{author}{Yan M},
  et~al.
\newblock \bibinfo{title}{Multi-objective optimization in cpu design space
  exploration: Attention is all you need}.
\newblock \bibinfo{journal}{arXiv preprint arXiv:2410.18368},
  \bibinfo{year}{2024}

\bibitem{Boom_Explorer_21_ICCAD}
\bibinfo{author}{Bai C}, \bibinfo{author}{Sun Q}, \bibinfo{author}{Zhai J},
  et~al.
\newblock \bibinfo{title}{{{BOOM-Explorer}}: {{RISC-V BOOM Microarchitecture
  Design Space Exploration Framework}}}.
\newblock In: \bibinfo{booktitle}{{{IEEE}}/{{ACM International Conference On
  Computer Aided Design}}, {{ICCAD}} 2021, {{Munich}}, {{Germany}},
  {{November}} 1-4, 2021}, \bibinfo{year}{2021}.
\newblock \bibinfo{pages}{1--9}

\bibitem{Mcpat-calib_TCAD_23}
\bibinfo{author}{Zhai J}, \bibinfo{author}{Bai C}, \bibinfo{author}{Zhu B},
  et~al.
\newblock \bibinfo{title}{{{McPAT-Calib}}: {{A RISC-V BOOM Microarchitecture
  Power Modeling Framework}}}.
\newblock \bibinfo{journal}{IEEE Transactions on Computer-Aided Design of
  Integrated Circuits and Systems}, \bibinfo{year}{2023}, \bibinfo{volume}{42}:
  \bibinfo{pages}{243--256}

\bibitem{MoEnDSE_GLSVLSI_23}
\bibinfo{author}{Wang D}, \bibinfo{author}{Yan M}, \bibinfo{author}{Teng Y},
  et~al.
\newblock \bibinfo{title}{A {{High-accurate Multi-objective Ensemble
  Exploration Framework}} for {{Design Space}} of {{CPU Microarchitecture}}}.
\newblock In: \bibinfo{booktitle}{Proceedings of the {{Great Lakes Symposium}}
  on {{VLSI}} 2023, {{Knoxville}}, {{TN}}, {{USA}}, {{June}} 5\textendash 7},
  \bibinfo{year}{2023}.
\newblock \bibinfo{pages}{379–383}

\bibitem{codes16_TrDSE}
\bibinfo{author}{Li D}, \bibinfo{author}{Wang S}, \bibinfo{author}{Yao S},
  et~al.
\newblock \bibinfo{title}{Efficient design space exploration by knowledge
  transfer}.
\newblock In: \bibinfo{booktitle}{Proceedings of the {{Eleventh
  IEEE}}/{{ACM}}/{{IFIP International Conference}} on {{Hardware}}/{{Software
  Codesign}} and {{System Synthesis}}, {{CODES}} 2016, {{Pittsburgh}},
  {{Pennsylvania}}, {{USA}}, {{October}} 1-7, 2016}, \bibinfo{year}{2016}.
\newblock \bibinfo{pages}{12:1--12:10}

\bibitem{iccad17_TrEE}
\bibinfo{author}{Li D}, \bibinfo{author}{Yao S}, \bibinfo{author}{Wang S},
  et~al.
\newblock \bibinfo{title}{Cross-program design space exploration by ensemble
  transfer learning}.
\newblock In: \bibinfo{editor}{S~Parameswaran}, ed., \bibinfo{booktitle}{2017
  {{IEEE}}/{{ACM International Conference}} on {{Computer-Aided Design}},
  {{ICCAD}} 2017, {{Irvine}}, {{CA}}, {{USA}}, {{November}} 13-16, 2017},
  \bibinfo{year}{2017}.
\newblock \bibinfo{pages}{201--208}

\bibitem{MPGM_ISAC_19}
\bibinfo{author}{Ding Y}, \bibinfo{author}{Mishra N}, \bibinfo{author}{Hoffmann
  H}.
\newblock \bibinfo{title}{Generative and multi-phase learning for computer
  systems optimization}.
\newblock In: \bibinfo{editor}{S~B Manne}, \bibinfo{editor}{H~C Hunter},
  \bibinfo{editor}{E~R Altman}, eds., \bibinfo{booktitle}{Proceedings of the
  46th {{International Symposium}} on {{Computer Architecture}}, {{ISCA}} 2019,
  {{Phoenix}}, {{AZ}}, {{USA}}, {{June}} 22-26, 2019}, \bibinfo{year}{2019}.
\newblock \bibinfo{pages}{39--52}

\bibitem{TrEnDSE_ICCAD_23}
\bibinfo{author}{Wang D}, \bibinfo{author}{Yan M}, \bibinfo{author}{Teng Y},
  et~al.
\newblock \bibinfo{title}{A transfer learning framework for high-accurate
  cross-workload design space exploration of cpu}.
\newblock In: \bibinfo{booktitle}{2023 {{IEEE}}/{{ACM International Conference
  On Computer Aided Design}} ({{ICCAD}})}, \bibinfo{year}{2023}

\bibitem{MetaDSE}
\bibinfo{author}{Xue R}, \bibinfo{author}{Wu H}, \bibinfo{author}{Yan M},
  et~al.
\newblock \bibinfo{title}{Metadse: A few-shot meta-learning framework for
  cross-workload cpu design space exploration}.
\newblock In: \bibinfo{booktitle}{{{DAC}}'25: 62nd {{ACM}}/{{IEEE Design
  Automation Conference}}, {{San Francisco}}, {{California}}, {{USA}}, {{July}}
  22 - 25, 2025}, \bibinfo{year}{2025}

\bibitem{eyerman_GA_2level_DATE_06}
\bibinfo{author}{Eyerman S}, \bibinfo{author}{Eeckhout L},
  \bibinfo{author}{De~Bosschere K}.
\newblock \bibinfo{title}{Efficient {{Design Space Exploration}} of {{High
  Performance Embedded Out-of-Order Processors}}}.
\newblock In: \bibinfo{booktitle}{Proceedings of the {{Design Automation}} \&
  {{Test}} in {{Europe Conference}}}, \bibinfo{year}{2006},
  volume~\bibinfo{volume}{1}.
\newblock \bibinfo{pages}{1--6}

\bibitem{mariani_meta_model_2009}
\bibinfo{author}{Mariani G}, \bibinfo{author}{Palermo G},
  \bibinfo{author}{Silvano C}, et~al.
\newblock \bibinfo{title}{Meta-model {{Assisted Optimization}} for {{Design
  Space Exploration}} of {{Multi-Processor Systems-on-Chip}}}.
\newblock In: \bibinfo{editor}{A~N{\'u}{\~n}ez}, \bibinfo{editor}{P~P
  Carballo}, eds., \bibinfo{booktitle}{12th {{Euromicro Conference}} on
  {{Digital System Design}}, {{Architectures}}, {{Methods}} and {{Tools}},
  {{DSD}} 2009, 27-29 {{August}} 2009, {{Patras}}, {{Greece}}},
  \bibinfo{year}{2009}.
\newblock \bibinfo{pages}{383--389}

\bibitem{ML-NSGA-II_09}
\bibinfo{author}{Mariani G}, \bibinfo{author}{Palermo G},
  \bibinfo{author}{Silvano C}, et~al.
\newblock \bibinfo{title}{Multi-processor system-on-chip {{Design Space
  Exploration}} based on multi-level modeling techniques}.
\newblock In: \bibinfo{editor}{W~A Najjar}, \bibinfo{editor}{M~J Schulte},
  eds., \bibinfo{booktitle}{Proceedings of the 2009 {{International
  Conference}} on {{Embedded Computer Systems}}: {{Architectures}},
  {{Modeling}} and {{Simulation}} ({{IC-SAMOS}} 2009), {{Samos}}, {{Greece}},
  {{July}} 20-23, 2009}, \bibinfo{year}{2009}.
\newblock \bibinfo{pages}{118--124}

\bibitem{NGSA_ANN_ATECS_13}
\bibinfo{author}{Mariani G}, \bibinfo{author}{Palermo G},
  \bibinfo{author}{Zaccaria V}, et~al.
\newblock \bibinfo{title}{Design-{{Space Exploration}} and {{Runtime Resource
  Management}} for {{Multicores}}}.
\newblock \bibinfo{journal}{ACM Trans. Embed. Comput. Syst.},
  \bibinfo{year}{2013}, \bibinfo{volume}{13}

\bibitem{ACOSSO_ASPDAC_15}
\bibinfo{author}{Wang H}, \bibinfo{author}{Zhu Z}, \bibinfo{author}{Shi J},
  et~al.
\newblock \bibinfo{title}{An accurate {{ACOSSO}} metamodeling technique for
  processor architecture design space exploration}.
\newblock In: \bibinfo{booktitle}{The 20th {{Asia}} and {{South Pacific Design
  Automation Conference}}}, \bibinfo{year}{2015}.
\newblock \bibinfo{pages}{689--694}

\bibitem{Multicube_ARCS_10}
\bibinfo{author}{Zaccaria V}, \bibinfo{author}{Palermo G},
  \bibinfo{author}{Castro F}, et~al.
\newblock \bibinfo{title}{Multicube {{Explorer}}: {{An Open Source Framework}}
  for {{Design Space Exploration}} of {{Chip Multi-Processors}}}.
\newblock In: \bibinfo{editor}{M~Beigl}, \bibinfo{editor}{F~J
  {Cazorla-Almeida}}, eds., \bibinfo{booktitle}{{{ARCS}} '10 - 23th
  {{International Conference}} on {{Architecture}} of {{Computing Systens}}
  2010, {{Workshop Proceedings}}, {{February}} 22-23, 2010, {{Hannover}},
  {{Germany}}}, \bibinfo{year}{2010}.
\newblock \bibinfo{pages}{325--331}

\bibitem{DSE_critical_PACT_10}
\bibinfo{author}{Navada S}, \bibinfo{author}{Choudhary N~K},
  \bibinfo{author}{Rotenberg E}.
\newblock \bibinfo{title}{Criticality-driven superscalar design space
  exploration}.
\newblock In: \bibinfo{booktitle}{Proceedings of the 19th International
  Conference on {{Parallel}} Architectures and Compilation Techniques},
  \bibinfo{year}{2010}.
\newblock \bibinfo{pages}{261--272}

\bibitem{mariani_MOSA_08}
\bibinfo{author}{Mariani G}, \bibinfo{author}{Palermo G},
  \bibinfo{author}{Silvano C}, et~al.
\newblock \bibinfo{title}{An {{Efficient Design Space Exploration Methodology}}
  for {{Multi-Cluster VLIW Architectures}} based on {{Artificial Neural
  Networks}}}.
\newblock \bibinfo{journal}{In: Proc. IFIP International Conference on Very
  Large Scale Integration VLSI - SoC}, \bibinfo{year}{2008}

\bibitem{mariani_correlation_based_DAC_10}
\bibinfo{author}{Mariani G}, \bibinfo{author}{Brankovic A},
  \bibinfo{author}{Palermo G}, et~al.
\newblock \bibinfo{title}{A correlation-based design space exploration
  methodology for multi-processor systems-on-chip}.
\newblock In: \bibinfo{editor}{S~S Sapatnekar}, ed.,
  \bibinfo{booktitle}{Proceedings of the 47th {{Design Automation Conference}},
  {{DAC}} 2010, {{Anaheim}}, {{California}}, {{USA}}, {{July}} 13-18, 2010},
  \bibinfo{year}{2010}.
\newblock \bibinfo{pages}{120--125}

\bibitem{mariani_OSCAR_TCAD_12}
\bibinfo{author}{Mariani G}, \bibinfo{author}{Palermo G},
  \bibinfo{author}{Zaccaria V}, et~al.
\newblock \bibinfo{title}{{{OSCAR}}: {{An Optimization Methodology Exploiting
  Spatial Correlation}} in {{Multicore Design Spaces}}}.
\newblock \bibinfo{journal}{IEEE Trans. Comput. Aided Des. Integr. Circuits
  Syst.}, \bibinfo{year}{2012}, \bibinfo{volume}{31}: \bibinfo{pages}{740--753}

\bibitem{mei2021zigzag}
\bibinfo{author}{Mei L}, \bibinfo{author}{Houshmand P}, \bibinfo{author}{Jain
  V}, et~al.
\newblock \bibinfo{title}{Zigzag: Enlarging joint architecture-mapping design
  space exploration for dnn accelerators}.
\newblock \bibinfo{journal}{IEEE Transactions on Computers},
  \bibinfo{year}{2021}, \bibinfo{volume}{70}: \bibinfo{pages}{1160--1174}

\bibitem{hong2023dosa}
\bibinfo{author}{Hong C}, \bibinfo{author}{Huang Q}, \bibinfo{author}{Dinh G},
  et~al.
\newblock \bibinfo{title}{Dosa: Differentiable model-based one-loop search for
  dnn accelerators}.
\newblock In: \bibinfo{booktitle}{Proceedings of the 56th Annual IEEE/ACM
  International Symposium on Microarchitecture}, \bibinfo{year}{2023}.
\newblock \bibinfo{pages}{209--224}

\bibitem{cong2018polysa}
\bibinfo{author}{Cong J}, \bibinfo{author}{Wang J}.
\newblock \bibinfo{title}{Polysa: Polyhedral-based systolic array
  auto-compilation}.
\newblock In: \bibinfo{booktitle}{2018 IEEE/ACM International Conference on
  Computer-Aided Design (ICCAD)}, \bibinfo{year}{2018}.
\newblock \bibinfo{pages}{1--8}

\bibitem{wang2021autosa}
\bibinfo{author}{Wang J}, \bibinfo{author}{Guo L}, \bibinfo{author}{Cong J}.
\newblock \bibinfo{title}{Autosa: A polyhedral compiler for high-performance
  systolic arrays on fpga}.
\newblock In: \bibinfo{booktitle}{The 2021 ACM/SIGDA International Symposium on
  Field-Programmable Gate Arrays}, \bibinfo{year}{2021}.
\newblock \bibinfo{pages}{93--104}

\bibitem{shao2014aladdin}
\bibinfo{author}{Shao Y~S}, \bibinfo{author}{Reagen B}, \bibinfo{author}{Wei
  G~Y}, et~al.
\newblock \bibinfo{title}{Aladdin: A pre-rtl, power-performance accelerator
  simulator enabling large design space exploration of customized
  architectures}.
\newblock \bibinfo{journal}{ACM SIGARCH Computer Architecture News},
  \bibinfo{year}{2014}, \bibinfo{volume}{42}: \bibinfo{pages}{97--108}

\bibitem{Chip-Chat23}
\bibinfo{author}{Blocklove J}, \bibinfo{author}{Garg S}, \bibinfo{author}{Karri
  R}, et~al.
\newblock \bibinfo{title}{Chip-chat: Challenges and opportunities in
  conversational hardware design}.
\newblock In: \bibinfo{booktitle}{5th {ACM/IEEE} Workshop on Machine Learning
  for CAD, {MLCAD} 2023, Snowbird, UT, USA, September 10-13, 2023},
  \bibinfo{year}{2023}.
\newblock \bibinfo{pages}{1--6}.
\newblock \urlprefix\url{https://doi.org/10.1109/MLCAD58807.2023.10299874}

\bibitem{AutoChip23}
\bibinfo{author}{Thakur S}, \bibinfo{author}{Blocklove J},
  \bibinfo{author}{Pearce H}, et~al.
\newblock \bibinfo{title}{Autochip: Automating {HDL} generation using {LLM}
  feedback}.
\newblock \bibinfo{journal}{CoRR}, \bibinfo{year}{2023},
  \bibinfo{volume}{abs/2311.04887}.
\newblock \urlprefix\url{https://doi.org/10.48550/arXiv.2311.04887}

\bibitem{DeLorenzo24}
\bibinfo{author}{DeLorenzo M}, \bibinfo{author}{Chowdhury A~B},
  \bibinfo{author}{Gohil V}, et~al.
\newblock \bibinfo{title}{Make every move count: Llm-based high-quality {RTL}
  code generation using {MCTS}}.
\newblock \bibinfo{journal}{CoRR}, \bibinfo{year}{2024},
  \bibinfo{volume}{abs/2402.03289}.
\newblock \urlprefix\url{https://doi.org/10.48550/arXiv.2402.03289}

\bibitem{OriGen24}
\bibinfo{author}{Cui F}, \bibinfo{author}{Yin C}, \bibinfo{author}{Zhou K},
  et~al.
\newblock \bibinfo{title}{Origen:enhancing {RTL} code generation with
  code-to-code augmentation and self-reflection}.
\newblock \bibinfo{journal}{CoRR}, \bibinfo{year}{2024},
  \bibinfo{volume}{abs/2407.16237}.
\newblock \urlprefix\url{https://doi.org/10.48550/arXiv.2407.16237}

\bibitem{VeriGen23}
\bibinfo{author}{Thakur S}, \bibinfo{author}{Ahmad B}, \bibinfo{author}{Fan Z},
  et~al.
\newblock \bibinfo{title}{Benchmarking large language models for automated
  verilog {RTL} code generation}.
\newblock In: \bibinfo{booktitle}{Design, Automation {\&} Test in Europe
  Conference {\&} Exhibition, {DATE} 2023, Antwerp, Belgium, April 17-19,
  2023}, \bibinfo{year}{2023}.
\newblock \bibinfo{pages}{1--6}.
\newblock \urlprefix\url{https://doi.org/10.23919/DATE56975.2023.10137086}

\bibitem{VerilogEval23}
\bibinfo{author}{Liu M}, \bibinfo{author}{Pinckney N~R},
  \bibinfo{author}{Khailany B}, et~al.
\newblock \bibinfo{title}{Verilogeval: Evaluating large language models for
  verilog code generation}.
\newblock In: \bibinfo{booktitle}{{IEEE/ACM} International Conference on
  Computer Aided Design, {ICCAD} 2023, San Francisco, CA, USA, October 28 -
  Nov. 2, 2023}, \bibinfo{year}{2023}.
\newblock \bibinfo{pages}{1--8}.
\newblock \urlprefix\url{https://doi.org/10.1109/ICCAD57390.2023.10323812}

\bibitem{MEV-LLM24}
\bibinfo{author}{Nadimi B}, \bibinfo{author}{Zheng H}.
\newblock \bibinfo{title}{A multi-expert large language model architecture for
  verilog code generation}.
\newblock \bibinfo{journal}{CoRR}, \bibinfo{year}{2024},
  \bibinfo{volume}{abs/2404.08029}.
\newblock \urlprefix\url{https://doi.org/10.48550/arXiv.2404.08029}

\bibitem{CodeV24}
\bibinfo{author}{Zhao Y}, \bibinfo{author}{Huang D}, \bibinfo{author}{Li C},
  et~al.
\newblock \bibinfo{title}{Codev: Empowering llms for verilog generation through
  multi-level summarization}.
\newblock \bibinfo{journal}{CoRR}, \bibinfo{year}{2024},
  \bibinfo{volume}{abs/2407.10424}.
\newblock \urlprefix\url{https://doi.org/10.48550/arXiv.2407.10424}

\bibitem{RTLLM24}
\bibinfo{author}{Lu Y}, \bibinfo{author}{Liu S}, \bibinfo{author}{Zhang Q},
  et~al.
\newblock \bibinfo{title}{{RTLLM:} an open-source benchmark for design {RTL}
  generation with large language model}.
\newblock In: \bibinfo{booktitle}{Proceedings of the 29th Asia and South
  Pacific Design Automation Conference, {ASPDAC} 2024, Incheon, Korea, January
  22-25, 2024}, \bibinfo{year}{2024}.
\newblock \bibinfo{pages}{722--727}.
\newblock \urlprefix\url{https://doi.org/10.1109/ASP-DAC58780.2024.10473904}

\bibitem{ITERTL24}
\bibinfo{author}{Wu P}, \bibinfo{author}{Guo N}, \bibinfo{author}{Xiao X},
  et~al.
\newblock \bibinfo{title}{{ITERTL:} an iterative framework for fine-tuning llms
  for {RTL} code generation}.
\newblock \bibinfo{journal}{CoRR}, \bibinfo{year}{2024},
  \bibinfo{volume}{abs/2407.12022}.
\newblock \urlprefix\url{https://doi.org/10.48550/arXiv.2407.12022}

\bibitem{AIvril24}
\bibinfo{author}{ul~Islam M}, \bibinfo{author}{Sami H},
  \bibinfo{author}{Gaillardon P}, et~al.
\newblock \bibinfo{title}{Aivril: Ai-driven {RTL} generation with verification
  in-the-loop}.
\newblock \bibinfo{journal}{CoRR}, \bibinfo{year}{2024},
  \bibinfo{volume}{abs/2409.11411}.
\newblock \urlprefix\url{https://doi.org/10.48550/arXiv.2409.11411}

\bibitem{VerilogCoder25}
\bibinfo{author}{Ho C}, \bibinfo{author}{Ren H}, \bibinfo{author}{Khailany B}.
\newblock \bibinfo{title}{Verilogcoder: Autonomous verilog coding agents with
  graph-based planning and abstract syntax tree (ast)-based waveform tracing
  tool}.
\newblock In: \bibinfo{editor}{T~Walsh}, \bibinfo{editor}{J~Shah},
  \bibinfo{editor}{Z~Kolter}, eds., \bibinfo{booktitle}{AAAI-25, Sponsored by
  the Association for the Advancement of Artificial Intelligence, February 25 -
  March 4, 2025, Philadelphia, PA, {USA}}, \bibinfo{year}{2025}.
\newblock \bibinfo{pages}{300--307}.
\newblock \urlprefix\url{https://doi.org/10.1609/aaai.v39i1.32007}

\bibitem{MAGE24}
\bibinfo{author}{Zhao Y}, \bibinfo{author}{Zhang H}, \bibinfo{author}{Huang H},
  et~al.
\newblock \bibinfo{title}{{MAGE:} {A} multi-agent engine for automated {RTL}
  code generation}.
\newblock \bibinfo{journal}{CoRR}, \bibinfo{year}{2024},
  \bibinfo{volume}{abs/2412.07822}.
\newblock \urlprefix\url{https://doi.org/10.48550/arXiv.2412.07822}

\bibitem{githubGitHubMmtatdiffgaussianrasterization}
\bibinfo{title}{{G}it{H}ub - mmt-at/diff-gaussian-rasterization: {C} {K}ernel
  --- github.com}.
\newblock
  \bibinfo{howpublished}{\url{https://github.com/mmt-at/diff-gaussian-rasterization}}

\end{thebibliography}
